\documentclass{jfm}

\usepackage{graphicx}
\usepackage{amsmath} 
\usepackage{newtxtext}
\usepackage{newtxmath}
\usepackage{tabularx}
\usepackage{natbib}
\usepackage{hyperref}
\usepackage{cleveref}
\usepackage{cancel}
\usepackage{color}
\usepackage[normalem]{ulem}
\usepackage{setspace}
\usepackage{tikz}
\usepackage{float}
\usepackage{caption}
\usepackage{setspace}
\usepackage{subcaption}
\tikzset{
main node/.style={inner sep=0,outer sep=0},
label node/.style={inner sep=0,outer ysep=.2em,outer xsep=.4em,font=\scriptsize,overlay},
strike out/.style={shorten <=-.2em,shorten >=-.5em,overlay}
}

\hypersetup{
    colorlinks = true,
    urlcolor   = blue,
    citecolor  = black,
}

\newcommand{\RomanNumeralCaps}[1]
\linenumbers

\title{Impact of compressibility approximations on atmospheric gravity wave characteristics}

\author{Hrishikesh Sivanandan\aff{1}
  \corresp{\email{hrsiva@iu.edu}},
  Matthew Churchfield \aff{2},
  Ben Kravitz \aff{1},
  Eliot Quon \aff{3},
 \and  Mehtab Khan \aff{4}}

\affiliation{
\aff{1} Department of Earth and Atmospheric Sciences, Indiana University, Bloomington, Indiana, USA
\aff{2} National Wind Technology Center, National Laboratory of the Rockies, Golden, Colorado, USA 
\aff{3} Aeolus Labs, San Francisco, California, USA
\aff{4} Department of Flow Physics and Technology, Faculty of Aerospace Engineering, Delft University of Technology, The Netherlands
}

\begin{document}
\maketitle

\newcommand{\fix}{\textcolor{red}}
\newcommand{\fixed}{\textcolor{blue}}
\begin{abstract}
    Atmospheric gravity waves are a common phenomenon in the atmosphere and are of critical importance to understanding power generation from wind plants.  Common sources of atmospheric gravity waves are stratified flow over topography or wind deceleration ahead of wind farms.  Simulations to understand atmospheric boundary layer conditions and wind power generation often solve the Navier-Stokes equations in incompressible Boussinesq form, which allows for density variation as a function of potential temperature only; the purpose of this study is to understand how much that simplification matters.
    
    In this work, we evaluate simulated internal gravity wave characteristics by modeling stratified flow over bell-shaped hills using three forms of the governing equations: the linearized incompressible Boussinesq form, the full incompressible Boussinesq form, and the fully compressible form. While all formulations predict qualitatively similar wave structures, the compressible equations predict larger parcel displacements at higher altitudes, higher variations in apparent wavelengths, and more differences in wave angle, particularly at low Froude numbers. These differences can be attributed to compressibility and non-linear effects. The physical implications of differing wave characteristics become especially important in low Froude number regimes. These differences can, in turn, lead to differing predictions of wind farm performance and the characteristics of wind farm-induced gravity waves, underscoring the potential importance of compressibility effects in accurately capturing flow behavior in strongly stratified environments.
\end{abstract}



\section{Introduction}
\label{sec:intro}
Wind and solar energy are currently 15\% of power generation in the United States and increasing with calls for reductions in fossil fuel consumption and increased energy security \citep{IRENA_2025}. With this increase in renewable energy comes the challenge of maintaining a stable grid under variable generation. Thus, predicting wind power generation is of critical importance to energy security and grid resilience.

Annual energy production (AEP) estimates at prospective wind farm sites are often overestimated during the pre-construction phase \citep{Clifton_2016, Bleeg2018, Yan_2022}. This overestimation has been attributed to several factors, including wake interactions, atmospheric stability effects \citep{abkar_influence_2015}, complex terrain influences \citep{draxl_mountain_2020}, and limitations in turbulence modeling.

One potentially overlooked factor in energy yield estimation is the lack of consideration of atmospheric gravity wave (AGW) effects. AGWs are oscillations in a stably stratified atmospheric flow field, where buoyancy is the restoring force \citep{Rauber2015, nappo_book, Feng2025}. They are perturbed in the stably stratified layers of the atmosphere by various notable sources, including topography \citep{Lyra1943, queney_problem_1948, Scorer1949, Gossard1953, Lott1998, Smith2006}, convection \citep{Fritts2003}, severe weather \citep{Jordan1972}, wind shear \citep{Dong2023}, atmospheric jets and fronts \citep{Plungoven_2014}, coastal land-sea transition, and large wind farms. 
Further, they drive large-scale circulation in the middle atmosphere by accelerating mean flow through the convergence of the gravity wave momentum flux \citep{Fritts1984} and control the quasi-biennial oscillation \citep{Lindzen1968}. When AGWs break in the free atmosphere, they cause clear air turbulence \citep{Koch_2005}.

Numerical studies have shown that wind farm-induced AGWs can propagate upstream and modulate the inflow conditions experienced by multiple rows of turbines \citep{allaerts_boundary-layer_2017, abkar_influence_2015} and also downstream, interacting with the wake of wind turbines \citep{Mass_2023}. Despite this, AGWs are not normally considered in pre-construction energy assessments. Furthermore, AGW flow simulations themselves are highly sensitive to numerical boundary condition choices, particularly the configuration of Rayleigh damping layers, raising concerns that some simulated AGWs may be inaccurate because of poor boundary condition handling \citep{Kosovic_2025, Mehtab2024b}. This leads to our central motivation: the misrepresentation -- or complete omission -- of AGWs in both design phase flow modeling and operational analysis likely contributes to inaccurate estimation of energy yield in modern wind farms.

AGWs have been studied as mountain waves for many years using the linearized incompressible Boussinesq equations (linear wave theory)  \citep{Lyra1943, queney_problem_1948, Scorer1949, Gossard1953, Long1955}. More recently, they have become relevant in wind farm flows where wind farm-induced AGWs and their impact on wind turbine power production have been studied, again using the incompressible Boussinesq form of the equations  \citep{allaerts_large_2015, allaerts_gravity_2018, allaerts_sensitivity_2019, Lanzilao2022, Lanzilao_2024}. The Boussinesq approximation simplifies density effects such that buoyancy-driven motions are retained while acoustic waves are not. In wind farm simulations, computational domains are often made very tall to resolve gravity waves generated by the farm. However, when the flow is assumed to be incompressible, wave characteristics do not change with decreasing atmospheric pressure, possibly leading to misprediction of the impact of wind farm-induced AGWs.

Anecdotally, the wind energy community is divided on whether AGWs are important in estimates of wind power. One group argues that wind farm-induced AGWs may not impact wind power production as much as numerical studies suggest. They attribute this to an increased sensitivity of incompressible Boussinesq codes to boundary conditions; these simulations are sensitive to the configuration of Rayleigh damping layers used to prevent artificial wave reflections at domain boundaries \citep{allaerts_large_2015, abkar_large-eddy_2017, Mehtab2024b}. Also, even though numerical studies suggest that wind farm-induced AGWs could significantly affect wind farm performance, their actual impact on power production has yet to be confirmed through observational evidence \citep{Kosovic_2025}. However, another group sees the impact of these waves on wind farm generation as realistic estimates \citep{wu_flow_2017, Allaerts2014, allaerts_boundary-layer_2017, Lanzilao2022, Lanzilao_2024}. It is difficult to reconcile these two viewpoints without an understanding of how the Boussinesq approximation actually changes the predicted characteristics of AGWs.

In this study, we investigate how different approximations of the fully compressible governing equations affect AGWs and their characteristics. To explore this, we analyze the flow over a bell-shaped hill, a flow scenario studied since the 1940s \citep{Lyra1943, queney_problem_1948, Scorer1949}. Although hills are not wind farms, the flow over hills has similarities to the flow over a wind farm. Wind farms act as semi-permeable obstructions that redirect some incoming flow upward, triggering AGWs akin to hills \citep{allaerts_boundary-layer_2017, Mass_2023}. We pursue this investigation using three different codes that differ in the level of approximations of the governing equations they solve: the linearized incompressible Boussinesq equations, the nonlinear incompressible Boussinesq equations, and the nonlinear compressible equations. We focus only on internal gravity waves (IGWs) in the free atmosphere, leaving out other types that are triggered when a capping inversion exists between the atmospheric boundary layer and the free atmosphere aloft. Our primary goal is to determine how varying levels of approximations made to the governing equations influence IGW characteristics such as amplitude, wavelength, and wave angles.

This investigation is organized as follows. Section~\ref{sec:governingEquations} provides an overview of the governing equations under different assumptions, which leads into Section~\ref{sec:oscillatoryNature} that discusses the oscillatory nature of these equations. Next, we discuss the numerical models and simulation domain in Section~\ref{sec:numericalModels}.  This is followed by a description of the design of the numerical experiment in Section \ref{sec:designOfExperiments}.  We describe the metrics used to extract the wave characteristics in Section \ref{sec:qois}. Section~\ref{sec:results} compares the resultant wave fields produced by the different forms of the governing equations, and we end with conclusions in Section~\ref{sec:conclu}.

\section {Governing Equations}
\label{sec:governingEquations}
To quantify the impact of compressibility approximations on AGW characteristics, it is essential to first examine the governing equations themselves. These include the equations for conservation of mass (continuity), momentum, and energy (expressed through potential temperature). The \textbf{fully compressible set of equations} (Equations.~\ref{compr:momentum_original}--\ref{compr:energy_original}), which has the fewest approximations, can be expressed by the following three equations.

The momentum equation is given by 
\begin{equation}
\frac{D}{Dt}\left( \rho \mathbf{u} \right)= -\nabla p + \rho\mathbf{g} -\nabla \cdot \mathbf{\tau} + \mathbf{F_{ext}}
\label{compr:momentum_original}
\end{equation}
where $\rho$ represents density of the fluid, $\mathbf{u}$ is the three-dimensional velocity vector,  $p$ is the pressure, $\mathbf{g}$ is the acceleration due to gravity,  $\mathbf{\tau}$ is the viscous shear stress tensor ($\displaystyle{\mathbf{\tau} = -2\mu\sigma_{ij}}$; where $\mu$ is the dynamic viscosity and $\sigma_{ij}$ is the Kronecker delta function which is $1$ when $i=j$ and $0$ otherwise). $\mathbf{F_{ext}}$ are external forces, including the Rayleigh damping layer and the sponge layers (described below) that are added to damp out AGW reflections near domain boundaries.

The continuity equation is given by
\begin{equation}
\frac{\partial}{\partial t}\rho + \nabla \cdot (\rho \mathbf{u})= 0.
\label{compr:continuity}
\end{equation}

And finally, the energy equation given by the conservation of potential temperature is
\begin{equation}
    \frac{D}{D t}(\rho \theta) = \nabla \cdot (\rho\alpha_{\theta}\nabla\theta)
\label{compr:energy_original}
\end{equation}
where $\theta$ is potential temperature and $\alpha_{\theta}$ is the thermal diffusivity. Here, we do not include source terms that quantify the impact of moisture on potential temperature in the energy equations, as we limit our focus to dry air.

In many studies of atmospheric flow, the fully compressible equations are simplified. If the flow velocity is much lower than the speed of sound (\(\text{Mach} < 0.3\)), then assuming constant density simplifies the governing equations and allows for larger time steps without significant loss of accuracy. Buoyancy effects, which depend on variable density, are critical in atmospheric flows, though, and they can be included even while assuming incompressibility through Boussinesq's approximation of the buoyancy term.  This assumption yields the \textbf{incompressible Boussinesq set of equations}, where $\rho$ is assumed to be constant in space and time, which reduces the continuity equation to
\begin{equation}
    \nabla \cdot \mathbf{u} = 0.
    \label{incompr:continuity}
\end{equation}

The Boussinesq approximation simplifies the momentum equations such that density changes are ignored except in the buoyancy term. We split $\rho$ into background ($\overline{\rho}(z)$) and perturbation ($\rho'$) terms.  This simplifies the momentum equation from Equation.~\ref{compr:momentum_original} to 
\begin{equation}
\overline{\rho}\left[1+\frac{\rho'}{\overline{\rho}}\right]\frac{D}{Dt}\mathbf{u} = -\nabla (p) + \left[\overline{\rho}+\rho'\right]\mathbf{g}+\nabla\cdot\tau+F_{ext}
    \label{bouss}
\end{equation}

Atmospheric gravity waves have traditionally been analyzed under the assumption of an inviscid, irrotational flow \citep{Nappo2012, Taylor1931}. Therefore, we can neglect the shear stress term $\nabla \cdot \tau $.  Furthermore, in equation \ref{bouss}, if we assume that the perturbation terms are very small vis-a-vis the background value, then $ 1 + (\rho'/\overline{\rho}) \approx 1$ on the left-hand side, and dividing both sides by $\overline{\rho}$, equation \ref{bouss} reduces to
\begin{equation}
    \frac{D}{Dt}\mathbf{u} = -\frac{1}{\overline{\rho}}\nabla (p) + \left[1+\frac{\rho'}{\overline{\rho}}\right]\mathbf{g} + \frac{\mathbf{F_{ext}}}{\overline{\rho}}.
    \label{bouss_momentum_2}    
\end{equation}
where, $\displaystyle{\left(\rho' / \overline{\rho}\right)\mathbf{g}}$ is the deviation of the buoyancy force from hydrostatic. Boussinesq assumes small vertical excursions of the air parcels in which density changes predominantly depend upon potential temperature changes and not so much on pressure changes with altitude.  Therefore, the density perturbation $\rho'$ in the buoyancy term becomes a function of potential temperature only.

Consider the definition of the potential temperature,   
\begin{equation}
    \overline{\theta} = \overline{T} \left( \frac{p_{0}}{\overline{p}} \right)^{R/C_{p}}.
\end{equation}
Using the ideal gas law, $\overline{T} = \overline{p} / \left(R_{d}\overline{\rho}\right)$, and standard thermodynamic relations 
($C_{p}-C_{v}=R_{d}$ and $\gamma = C_{p}/C_{v}$), where $C_p$, $C_v$ are the specific heat capacities at constant pressure and volume respectively, and $R_d$ is the universal gas constant. This expression can be rewritten in terms of mean pressure and density as
\begin{equation}
    \overline{\theta} = \frac{1}{R_{d}} \, \overline{p}^{(C_{v}/C_{p})} \, \overline{\rho}^{-1} \, p_{0}^{(R_{d}/C_{p})}.
\end{equation}

Taking the natural logarithm of both sides yields
\begin{equation}
    \ln(\overline{\theta}) = \frac{C_{v}}{C_{p}} \ln(\overline{p}) - \ln(\overline{\rho}) + k,
\end{equation}
where $k$ is a constant. Introducing perturbations about the mean state and applying the approximation 
$\ln(1+\epsilon) \approx \epsilon$ for small $\epsilon$, the linearized form of the potential temperature equation becomes
\begin{equation}
    \frac{\theta'}{\overline{\theta}} = \frac{1}{\gamma}\frac{p'}{\overline{p}} - \frac{\rho'}{\overline{\rho}}.
\end{equation}

Rearranging gives the density perturbation as
\begin{equation}
    \rho' = \frac{p'\overline{\rho}}{\gamma\overline{p}} - \overline{\rho}\frac{\theta'}{\overline{\theta}}.
\end{equation}
Substituting $c_{s}^{2} = \gamma\overline{p}/\overline{\rho}$ yields
\begin{equation}
    \rho' = -\overline{\rho}\frac{\theta'}{\overline{\theta}} + \frac{p'}{c_{s}^{2}}.
\end{equation}

In the Boussinesq approximation, density fluctuations associated with pressure variations are typically small compared to those arising from temperature (or potential temperature); therefore, they can be scaled out assuming small vertical parcel excursions \citep{Bannon1996, Dutton_1969}. Neglecting the pressure term and multiplying by the gravity vector leads to 
\begin{equation}
    \frac{\rho'}{\overline{\rho}} \mathbf{g} \approx -\frac{\theta'}{\overline{\theta}} \mathbf{g},
\end{equation}
which is then used in the buoyancy term of \Cref{bouss_momentum_2} resulting in the final form of the momentum equation,
\begin{equation}
    \frac{D}{Dt}\mathbf{u} = -\frac{1}{\overline{\rho}}\nabla (p) + \left[1-\frac{\theta'}{\overline{\theta}}\right]\mathbf{g} + \frac{\mathbf{F_{ext}}}{\overline{\rho}}.
    \label{bouss_momentum_}    
\end{equation}
It is important to note that in making the Boussinesq approximation, density variations are assumed to be exclusively dependent on potential temperature variations. However, in the compressible case, density changes depend on pressure variations and potential temperature variations with altitude.

Furthermore, assuming incompressibility also simplifies the energy equation (Eqn.~\ref{compr:energy_original}) to
\begin{equation}
    \frac{D\theta}{Dt}=\nabla\cdot(\alpha_{\theta}\nabla\theta)
    \label{bouss_energy}
\end{equation}

A further simplification is to linearize these equations, leading to the \textbf{linearized, incompressible Boussinesq set of equations}. The linearized Boussinesq equations are used to study small perturbations about a steady background state. They enable analytical derivation of wave characteristics (through the Taylor-Goldstein equation), making them well-suited for understanding the fundamental dynamics of wave motion and providing a baseline for interpreting wave behavior of flow over a hill.

Similar to $p$ and $\rho$, velocity ($\mathbf{u}$) can be split into a background state which only varies vertically and a spatially and temporally varying perturbation:  $q(x,z,t) = q_0(z) + q'(x,z,t)$. (Without loss of generality, we assume two-dimensional flow in the $x$- and $z$-directions only.) Assuming a steady, uniform background flow ($u=\overline{U}$), the incompressible continuity equation simplifies to
\begin{equation}
\frac{\partial}{\partial x}u'+ \frac{\partial}{\partial z}w'=0
\label{eqn:linear_continuity}
\end{equation}

Further, splitting the pressure into a background and perturbation, assuming the background flow to be hydrostatic, where $\displaystyle \partial \overline{p} / \partial z = \overline{\rho}g$, and neglecting products of perturbation terms, the linearized $z$-momentum equation becomes
\begin{equation}
    \frac{\partial w^{\prime}}{\partial t} + \overline{U}\frac{\partial w^{\prime}}{\partial x} + \frac{1}{\overline{\rho}}\frac{\partial p'}{\partial z} -\frac{\theta'}{\overline{\theta}}g = 0
    \label{eq:2D_incomp_momZ}
\end{equation}

Furthermore, the linearized $x$-momentum equation becomes


\begin{equation}
    \frac{\partial u'}{\partial t} + \overline{U}\frac{\partial u'}{\partial x}+ w'\frac{\partial \overline{U}}{\partial z} + \frac{1}{\overline{\rho}}\frac{\partial p'}{\partial x} = 0
    \label{eq:2D_incomp_momX}
\end{equation}

Lastly, if we assume an incompressible system and also that the air parcel oscillations are small, then the vertical motion of the parcels is an adiabatic process. This reduces the energy equation to
\begin{equation}
    \frac{\partial \theta'}{\partial t}+w'\frac{\partial \overline{\theta}}{\partial z} = 0
    \label{eq:energy_incomp}
\end{equation}

We summarize all of these governing equation sets in Table~\ref{table_eqns}.

\begin{table}
\begin{center}
\def~{\hphantom{0}}
\begin{tabular}{m{0.15\textwidth}ccc}
\textbf{Formulation} & ~~~~\textbf{Momentum}~~~~~ & ~~~~\textbf{Continuity}~~~~ & ~~~~\textbf{Energy}~~~~ \\
Compressible & {$\!\begin{aligned}
\frac{D}{Dt}\left( \rho \mathbf{u} \right) =  -\nabla p + \rho\mathbf{g}  -\nabla \cdot \mathbf{\tau} + \mathbf{F_{ext}}
\end{aligned}$} & $\frac{\partial}{\partial t}\rho + \nabla \cdot (\rho \mathbf{u})= 0$ & $\frac{D}{D t}(\rho \theta) = \nabla \cdot (\rho\alpha_{\theta}\nabla\theta)$ \\
\\ 
\\ 
Incompressible Boussinesq  & {$\!\begin{aligned}\frac{D}{Dt}\mathbf{u} =  -\frac{1}{\overline{\rho}}\nabla (p) +\left[1+\frac{\rho'}{\overline{\rho}}\right]\mathbf{g} + \frac{\mathbf{F_{ext}}}{\overline{\rho}}
\end{aligned}$} 
& $\nabla \cdot \mathbf{u} = 0$
& $\displaystyle{\frac{D\theta}{Dt}=\nabla \cdot(\alpha_\theta \nabla \theta)}$ 
\\ 
\\
Linearized Incompressible Boussinesq  & {$\!\begin{aligned}
\frac{\partial u'}{\partial t} + \overline{U}\frac{\partial u'}{\partial x}+ w'\frac{\partial \overline{U}}{\partial z} + \frac{1}{\overline{\rho}}\frac{\partial p'}{\partial x} = 0 
\\
\left(\frac{\partial}{\partial t} + \overline{U}\frac{\partial }{\partial x}\right)w' + \frac{1}{\overline{\rho}}\frac{\partial p'}{\partial z} -\frac{\theta'}{\overline{\theta}}g = 0
\end{aligned}$}

& $\frac{\partial}{\partial x}u'+ \frac{\partial}{\partial z}w'=0$ & $\displaystyle{\frac{\partial \theta'}{\partial t} +w' \frac{\partial \overline{\theta}}{\partial z}= 0 }$\\
\end{tabular}
\caption{Summary of the governing equations for the compressible, incompressible Boussinesq, and linearized incompressible Boussinesq formulations.}
\label{table_eqns}
\end{center}
\end{table}

\section{The Oscillatory Nature of the Problem}
\label{sec:oscillatoryNature}

\begin{figure}
\begin{center}
\includegraphics[width=0.40\textwidth]{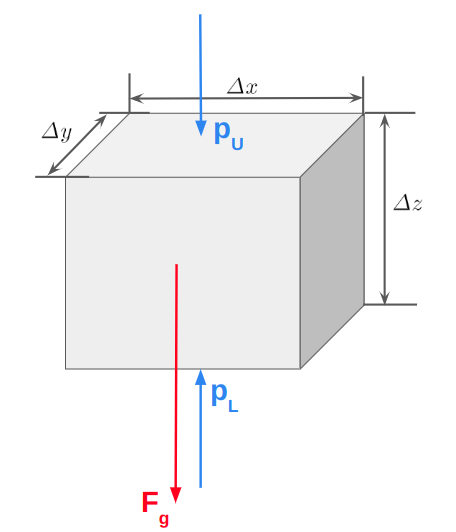}
\caption{An air parcel represented by a hexahedral volume 
of dimensions $\Delta x$, $\Delta y$, and $\Delta z$, and the pressure and gravitational force vectors acting upon it.}
\label{fig:air_parcel}
\end{center}
\end{figure} 

The dynamical system of a stably stratified fluid, described in the governing equations presented in Section \ref{sec:governingEquations}, is oscillatory.  Air parcels moving horizontally through a stably stratified atmosphere in which there is some vertical perturbation (e.g., topography) will undergo oscillations, and a wave field will result. To examine this behavior, we construct a simplified, one-dimensional, inviscid representation of the stably stratified atmosphere. This allows us to evaluate the parcel's vertical motion under different prescribed potential temperature profiles, $\theta(z)$, as described by the compressible and incompressible Boussinesq forms of the governing equations.

Consider an air parcel subject to pressure and gravitational forces, as shown in figure~\ref{fig:air_parcel}, that has a hexahedral volume of dimensions $\Delta x \times \Delta y \times \Delta z$ and mass $m$. Using Newton's second law, the vertical component of acceleration can be expressed as
\begin{equation}
\begin{aligned}
    \frac{\partial w}{\partial t} & = \frac{1}{m}\left(-p_U \Delta x\Delta y + p_L \Delta x\Delta y - F_g\right) \\
                                  & = \frac{1}{m}\left[-\left(\frac{p_U - p_L}{\Delta z}\right) \Delta V - F_g \right] \\
                                  & = -\frac{1}{\rho}\frac{\partial p}{\partial z} - \frac{F_g}{m},
\end{aligned}
\end{equation}
where $p_U$ and $p_L$ are the pressures on the upper and lower faces of the parcel, $\Delta V = \Delta x \Delta y \Delta z$ is the parcel volume, and $F_g$ is the gravitational force acting on the parcel. Also, $\partial p / \partial z = (p_U-p_L)/\Delta z$ becomes true as the parcel thickness $\Delta z$ becomes very small.  For the basic compressible formulation of the governing equations, the gravitational force is $mg$, which results in the vertical parcel acceleration,
\begin{equation}
    \frac{\partial w}{\partial t} = -\frac{1}{\rho}\frac{\partial p}{\partial z} + g.
    \label{eq:net_vert_acc}
\end{equation}
Note that in equation~\ref{eq:net_vert_acc}, we have switched the sign on the gravity term because going forward, we use the notation that $g$ is equal to $-9.81$ m/s$^2$. 

For the incompressible Boussinesq formulation, the density is assumed constant and denoted by $\rho_0$, which means the gravity term, without any modification, would be insensitive to buoyancy effects caused by temperature differences in the fluid.  Therefore, Boussinesq proposed that the gravity term be modified as $\left(1-\left[(\theta-\theta_{ref})/\theta_{ref}\right]\right)mg$, where $\theta$ is the potential temperature of the air parcel and $\theta_{ref}$ is the reference potential temperature.  This makes the gravity term sensitive to potential temperature differences in the fluid. The vertical acceleration of the air parcel in the incompressible Boussinesq form of the equations is
\begin{equation}
    \frac{\partial w}{\partial t} = - \frac{1}{\rho_0}\frac{\partial p}{\partial z} + \left[1- \frac{\theta-\theta_{ref}}{\theta_{ref}}\right]g.
    \label{eq:net_vert_acc_Bouss}
\end{equation}

In both equation~\ref{eq:net_vert_acc} and equation~\ref{eq:net_vert_acc_Bouss}, we assume the pressure gradient to be that of a background hydrostatic environment, which can be determined under the hydrostatic situation, $\partial w /\partial t  = 0$. We begin with the compressible form in equation~\ref{eq:net_vert_acc}, consider hydrostatic conditions, and solve for the pressure gradient, yielding
\begin{equation}
    \frac{\partial p}{\partial z} = \rho g.
    \label{eq:net_vert_acc_hydrostatic}
\end{equation}
We relate density to pressure and potential temperature using the ideal gas law, $p=\rho R T$, and the potential temperature relation, $\theta = T \; \left(p_0/p\right)^{R/C_p}$, where $R$ is the universal gas constant, $p_0$ is the surface pressure  and $C_p$ is the specific heat capacity of air at constant pressure. In this work, we only examine linear variations of potential temperature with height such that $\theta(z) = \theta_0+\Gamma z$, where $\theta_0$ is the surface temperature at $z=0$ and $\Gamma$ is the potential temperature lapse rate. Combining the ideal gas law with the potential temperature relation, we express density as 
\begin{equation}
    \rho = \frac{p_0^{\frac{R}{C_p}}}{R\;\,\left(\theta_0+\Gamma z \right)}p^{1-\frac{R}{C_p}}.
\end{equation}
Substituting this expression of density back into equation~\ref{eq:net_vert_acc_hydrostatic}, we find that
\begin{equation}
    \frac{\partial p}{\partial z} = \frac{p_0^{\frac{R}{C_p}}g}{R\;\,(\theta_0+\Gamma z)}p^{1-\frac{R}{C_p}},
\label{eq:PGF_comp_intermediate}
\end{equation} 
which is a separable differential equation that can be solved through the integral
\begin{equation}
    \int_{p_0}^{p} p^{\frac{R}{C_p}-1} dp = \frac{p_0^{\frac{R}{C_p}}g}{R}\int_{0}^{z} \frac{1}{\theta_0+ \Gamma z} \textrm{d}z,
\end{equation}
which has the solution for pressure,
\begin{equation}
      p(z) = \left[p_0^{\frac{R}{C_p}}\left(1+\frac{g}{\Gamma C_p}\ln\left(1+\frac{\Gamma z}{\theta_0}\right)\right)\right]^{\frac{C_p}{R}}.
\label{eq:p_comp}
\end{equation}
We can either differentiate equation~\ref{eq:p_comp} with respect to $z$ or insert it into equation~\ref{eq:PGF_comp_intermediate} to obtain the hydrostatic background pressure gradient term,
\begin{equation}
     \frac{\partial p}{\partial z} (z)=\frac{p_0^{\frac{R}{C_p}}g}{R \;\,(\theta_0+\Gamma z)}\left[p_0^{\frac{R}{C_p}}\left(1+\frac{g}{\Gamma C_p}\ln\left(1+\frac{\Gamma z}{\theta_0}\right)\right)\right]^{\frac{C_p}{R}-1}.
     \label{eq:PGF_comp}
\end{equation}

Solving for the hydrostatic vertical pressure gradient for the incompressible Boussinesq formulation is a simpler matter.  Under the hydrostatic conditions, starting with equation~\ref{eq:net_vert_acc_Bouss}, we find that pressure gradient is
\begin{equation}
    \frac{\partial p}{\partial z}(z)  = \rho_0\left(1-\frac{\theta_0+\Gamma z-\theta_{ref}}{\theta_{ref}}\right) g.
    \label{eqn:PGF_Bouss}
\end{equation}
We can also obtain pressure as a function of height in the hydrostatic incompressible Boussinesq formulation by integrating equation~\ref{eqn:PGF_Bouss} with height with the result
\begin{equation}
    p(z) = p_0 +\frac{\rho_0}{\theta_{ref}}\left[(2\theta_{ref}-\theta_0)z+\frac{\Gamma z^2}{2}\right]g.
\end{equation}

We now have everything to solve for the vertical motion of the air parcel. Before we begin, it is important to understand a few points. First, we denote the neutral height from which the parcel is perturbed as $z_n$ and the positive and negative perturbations from that height as $\pm \delta z$. Also, the pressure and pressure gradient that the parcel experiences is the background atmospheric values that we solved for using the hydrostatic condition. Lastly, as the parcel moves, it retains the potential temperature it had at its neutral height because the process is assumed adiabatic. For the compressible case, conservation of potential temperature does not imply that the parcel temperature remains constant. Although, by definition, potential temperature is conserved under adiabatic conditions, the parcel’s absolute temperature and density evolve as it equilibrates to the ambient pressure at its displaced height. Vertical displacements induce parcel expansion or compression, resulting in work exchange with the surrounding environment and, consequently, a change in temperature. This time we begin with the incompressible Boussinesq form and move to the compressible form.  We start by substituting the solution for the background pressure gradient, equation~\ref{eqn:PGF_Bouss} into equation~\ref{eq:net_vert_acc_Bouss} for vertical motion to get
\begin{equation}
\begin{aligned}
    \frac{\partial w}{\partial t} &= - \frac{1}{\rho_0}\frac{\partial p}{\partial z} + \left(1- \frac{\theta-\theta_{ref}}{\theta_{ref}}\right)g \\
    &= -\left(1-\frac{\theta_0+\Gamma(z_n+\delta z)-\theta_{ref}}{\theta_{ref}}\right)g+\left(1-\frac{\theta_0+\Gamma z_n-\theta_{ref}}{\theta_{ref}}\right)g \\
                                  &= \left(\frac{\Gamma g}{\theta_{ref}}\right) \delta z.
    \label{eq:bouss_acceleration}
\end{aligned}
\end{equation}

We do the same for the compressible case, starting from equation~\ref{eq:net_vert_acc}. We need to know the density of the parcel at any perturbation height about the neutral height. We use the ideal gas law while recognizing that the parcel keeps its potential temperature from the neutral height, but has the pressure of the background environment at the perturbed height. From this, the density of the air parcel at any height is
\begin{equation}
    \rho(z_n+\delta z) = \frac{p_0^{\frac{R}{C_p}}}{R \;\, (\theta_0+\Gamma z_n)}\left[p(z_n+\delta z)\right]^{1-\frac{R}{C_p}}.
    \label{eq:p_pert_comp}
\end{equation}
Also, the pressure gradient at the perturbation height using equation~\ref{eq:PGF_comp} is
\begin{equation}
    \frac{\partial p}{\partial z}(z_n+\delta z) = \frac{p_0^{\frac{R}{C_p}}g}{R \;\,[\theta_0+\Gamma (z_n+\delta z)]}\left[p(z_n+\delta z)\right]^{1-\frac{R}{C_p}}.
    \label{eq:PGF_pert_comp}
\end{equation}
Substituting equation~\ref{eq:p_pert_comp} and equation~\ref{eq:PGF_pert_comp} into equation~\ref{eq:net_vert_acc}, we find that
\begin{equation}
\begin{aligned}
    \frac{\partial w}{\partial t} &= -\frac{1}{\rho} \frac{\partial p}{\partial z} + g \\
    &= -\frac{R\;\,(\theta_0+\Gamma z_n)}{p_0^{\frac{R}{C_p}}\left[p(z_n+\delta z)\right]^{1-\frac{R}{C_p}}}\frac{p_0^{\frac{R}{C_p}}\left[p(z_n+\delta z)\right]^{1-\frac{R}{C_p}}}{R \;\,[\theta_0+\Gamma(z_n+\delta z)]}+g \\
    &= \left(1-\frac{\theta_0+\Gamma z_n}{\theta_0+\Gamma(z_n+\delta z)}\right)g.
    \label{eq:comp_acceleration}
\end{aligned}
\end{equation}

In equation~\ref{eq:bouss_acceleration}, the equation for parcel motion in the incompressible Boussinesq form of the equations, we recognise that $N = \sqrt{\Gamma |g| /\theta_{ref}}$, where $N$ is the buoyancy or Brunt-Väisälä frequency. Thus, equation~\ref{eq:bouss_acceleration} can be written as
\begin{equation}
    \frac{\partial^2 z}{\partial t^2} = -N^2 (z-z_n).
\end{equation}
The oscillatory solution to this equation is $z(t) = z_n + z_0 \cos(Nt)$, where $z_0$ is the initial position. This is analogous to a mass-spring dynamical system. Generally, $\theta_{ref}$ is set to 300 K and sometimes it is set to the initial temperature at the surface. Because $\Gamma$ is constant, $N$ takes a single constant value which implies that the oscillation frequency is the same at all heights.

Next, we investigate equation~\ref{eq:comp_acceleration}, which too represents an oscillatory system, but one in which the oscillation frequency is nonconstant in $\delta z$. It resembles the oscillation of a mass attached to a nonlinear spring. We can better understand this behavior by linearizing about $z_n$, which yields
\begin{equation}
    \frac{\partial w}{\partial t} = \frac{\Gamma}{\theta_0+\Gamma z_n}g
    \label{eq:comp_acc}
\end{equation}
If we set $\theta_{ref} = \theta_0+\Gamma z_n$ in the incompressible Boussinesq form, we get the same result as in the linearization of the compressible form.  The linearized result shows that in the compressible framework, the buoyancy frequency varies with height and can be defined as 
\begin{equation}
    N(z) = \sqrt{\frac{\Gamma |g|}{\theta_0+\Gamma z}}
    \label{eq:buoy_freq_comp}
\end{equation}
This result is remarkably similar to the incompressible Boussinesq result in equation~\ref{eq:bouss_acceleration}. In fact, it is the same if $\theta_{ref}=\theta_0+\Gamma z $, the temperature at the reference height.

\begin{figure}
    \centering
        \includegraphics[width=0.9\textwidth]{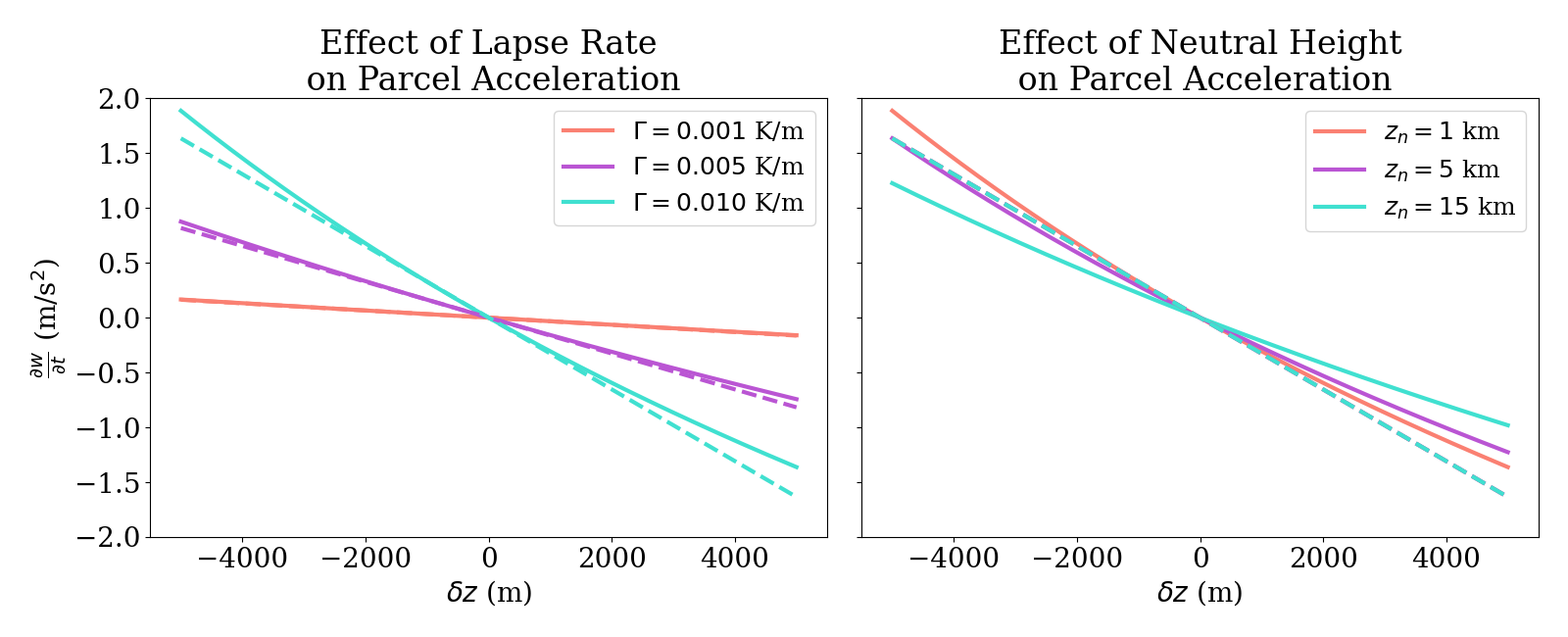}

    \caption{Parcel acceleration as a function of vertical displacement for varying lapse rate at a fixed $z_n = 1$ km (left) and for varying neutral height at a fixed $\Gamma =  0.01$ K$/$km (right). The solid lines represent the compressible formulation, while the dashed lines represent the incompressible Boussinesq formulation.}
    \label{fig:buoy_freq_z}
\end{figure}

The left plot in Figure \ref{fig:buoy_freq_z} shows the parcel acceleration as a function of the perturbation height, $\delta z$, away from the neutral height, $z_n$. This is the vertical acceleration that the parcel experiences at any height $\delta z$ along its oscillatory path. The plot on the left shows how acceleration varies with the lapse rate and the right plot shows how it varies with the neutral height about which the oscillation takes place. The solid lines are the results for the compressible flow equations and the dashed lines are that of the incompressible Boussinesq equations.

We notice a few important things. First, the incompressible Boussinesq behavior is linear, whereas the compressible behavior is slightly nonlinear. In the left plot, we see that as the lapse rate increases, the compressible behavior becomes more nonlinear. For the smallest lapse rate, the compressible and incompressible Bousinesq results match near perfectly. For a lapse rate of 0.005 $K/m$, the match is still very good even for a parcel displacement of $\pm 5$ km. Second, we notice that in the compressible formulation, the parcel acceleration depends on the neutral height. The greater the neutral height, the greater the acceleration magnitude for a given parcel displacement. The incompressible Boussinesq formulation has no dependence on parcel neutral height, however (all three lines are stacked one above the other, so we only see $z_n =$ 15 km line). This is due to the definition of $\theta_{ref}$ in the buoyancy term. If $\theta_{ref}$ were made height dependent, the incompressible Boussinesq formulation would behave more like the compressible formulation. In fact, according to equation \ref{eq:comp_acc}, if $\theta$ were defined as $\theta_0 + \Gamma z_n$, i.e. the ``background'' or ``base-state'' potential temperature, incompressible Boussinesq would be the linearized form of the compressible formulation. 

These differences highlight an important limitation of the incompressible Boussinesq approximation of the compressible equation set: it is not meant to be used for flows with large vertical parcel motion. Looking at the left plot of figure~\ref{fig:buoy_freq_z}, we can see that for typical lapse rates and for vertical motions around $\pm$2 km, the incompressible Boussinesq and compressible formulations match quite well. Looking at the right plot of figure~\ref{fig:buoy_freq_z}, we also see that the incompressible Boussinesq formulation matches the compressible formulation better at lower neutral heights. This is simply because we chose $z_{ref}$ in the incompressible Boussinesq formulation as 300 K which is the same as the temperature at the surface. The main point, though, is that the incompressible Boussinesq formulation becomes less valid as the ``base-state'' temperature deviates more from the $\theta_{ref}$. For simulations with taller domains with more temperature variation, we will see that incompressible Boussinesq deviates from the compressible results more than for shorter domains with less temperature variation.

\section{Numerical Models}
\label{sec:numericalModels}

\subsection{Solvers}
\label{subsec:solvers}
We use three different codes to solve the three forms of the governing equations outlined in Table \ref{table_eqns} applied to gravity waves formed over topography.

We solve the compressible, non-hydrostatic form of the equations using the Energy Research and Forecasting (ERF) code \citep{ERF_JOSS}. It solves the equations in terrain-following, non-hydrostatic pressure coordinates using a third-order accurate Runge-Kutta scheme for time integration and third-order accurate spatial discretization schemes for the advection terms for momentum and scalars.  ERF is built upon AMReX, a software framework for block-structured adaptive mesh refinement (AMR) \citep{AMReX_JOSS,AMReX_IJHPCA1}.

The incompressible Boussinesq form of the equations is solved with the Simulator fOr Wind Farm Applications (SOWFA) \citep{churchfield_2012}.  It is built upon the OpenFOAM computational fluid dynamics toolbox \citep{OpenFOAM,greenshieldsweller2022} and employs spatially and temporally second-order accurate finite-volume numerical schemes.  The equations are solved sequentially using a predictor-corrector method.  Although the code handles unstructured meshes, we employ hexahedral, terrain-following meshes in this study.

The linearized, incompressible Boussinesq form of the equations are solved using the Linear Buoyancy Wave (LBoW) simulator \citep{LBoW_Allaerts_2022} that provides a semi-analytical solution to the Taylor-Goldstein relation \citep{Taylor1931,Goldstein1931}. The Taylor-Goldstein relation is a wave equation for linear gravity waves in an irrotational, frictionless, adiabatic flow. LBoW solves the equation semi-analytically using the fast Fourier transform algorithm.

To ensure a fair comparison between all three models, surface friction is not included. This does not detract from the realism of the simulations because we are analyzing waves in the free atmosphere where frictional effects are not dominant. We also do not include the effect of Coriolis forces, although ERF and SOWFA are capable of including this effect.  The effect of Coriolis on gravity waves is a second-order effect that could be included in future studies.

\subsection{Simulation Domain}
\label{subsec:domain}

\begin{figure}
\begin{center}
\includegraphics[width=\textwidth]{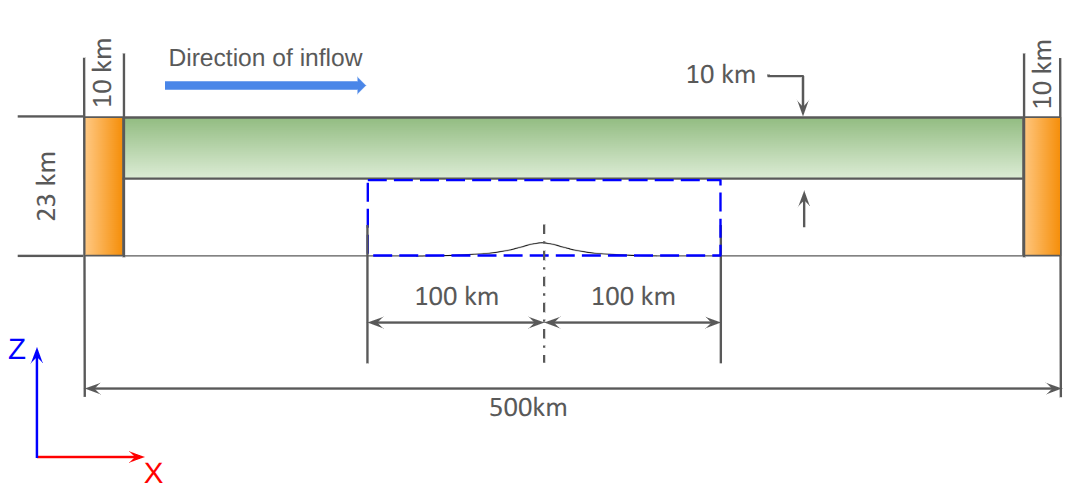}
\caption{An overview of the simulation domain. The top green region represents a 10-km thick damping layer used to absorb downward-propagating reflected waves, while the 10-km thick orange regions at the inlet and outlet denote damping layers that minimize lateral wave reflection.}
\label{fig:setup}
\end{center}
\end{figure}

We investigate the two-dimensional problem of atmospheric gravity waves generated by a bell-shaped hill subject to a uniform, non-turbulent, inviscid, stably-stratified inflow as shown in figure~\ref{fig:setup}.  The hill height is defined by the ``Witch of Agnesi'' function 
\begin{equation}
    h(x) = \frac{h_{max}}{1+\left( \frac{x}{L} \right)^2},
\end{equation}
where $h_{max}$ is the hill maximum height and $L$ is the hill half width at $h_{max}/2$.  Flow is from the left to right and the hill is in the middle of the domain at the bottom. The $x$-coordinate is in the flow direction and the $z$-coordinate is in the vertical.  The overall domain is 500 km long and 23 km tall.  In ERF and SOWFA, the left boundary of the domain is specified inflow, and the right is outflow.  The lower boundary is a slip wall; the upper boundary is also a slip wall, but with a damping layer to mitigate spurious wave reflections.  We also apply damping layers to the left and right boundaries for the same reason.  We discuss the details of the damping below in section \ref{subsec:damping}.  In LBoW, the domain is periodic in the horizontal, but since we use a 500 km long domain, we minimize any mismatch between the periodic LBoW and non-periodic ERF and SOWFA calculations.  The upper boundary condition in LBoW is radiative.

ERF and SOWFA use terrain-conformal meshes of 200-m horizontal resolution and 50-m vertical resolution.  In ERF, mesh deformation is only in the vertical, which means that vertical grid lines remain perfectly vertical and there is some mesh non-orthogonality over the hill.  In SOWFA, mesh deformation is in all three directions meaning that vertical grid lines project normal from the terrain before gradually becoming vertical to ensure a high degree of mesh orthogonality.  LBoW is different than ERF and SOWFA, which are computational fluid dynamics solvers, so the LBoW mesh is simply a Cartesian array of points that does not explicitly deform to the terrain---the effect of terrain is implicit in the bottom boundary condition applied to the solution to the Taylor-Goldstein equation. The LBoW meshes are of 200-m horizontal resolution and 50-m vertical resolution.

\subsection{Damping Layers for Mitigating Spurious Reflections}
\label{subsec:damping}
In the ERF and SOWFA simulations, damping layers are used to mitigate spurious wave reflections at the upper, inflow, and outflow computational domain boundaries.  The use of damping layers in calculations involving atmospheric gravity waves is well documented \citep{allaerts_large_2015,Lanzilao2022,Lanzilao_2024,Mehtab2024b}. They are implemented as artificial forcing terms on the right-hand side of the momentum equations, where the velocity is gradually relaxed to a reference state throughout a finite-thickness layer adjacent to the domain boundary. A general expression of the damping term is 
\begin{equation}
    \mathbf{F} = -\frac{1}{\tau} f(d) \; \textbf{c} \cdot(\textbf{u}-\textbf{u}_{ref})
\label{eq:damping}
\end{equation}
where $\tau$ is the damping time scale, $f(d)$ is the strength function of a specified shape, $d$ is the normal distance from the boundary toward the inside of the domain, $\textbf{c}$ is the vector that selects which components to damp, $\textbf{u}$ is the damped quantity, either momentum or velocity, and $\textbf{u}_{ref}$ is the reference value. 

There are various choices of strength function, $f(d)$.  We use the default functions in ERF and implemented them into SOWFA.  The strength function on the top boundary damping layer is a half period of a squared sin function which smoothly blends from one at the domain boundary to zero at the inside edge of the damping layer inside the domain, $f(d) = \sin^2[0.5 \pi (1-d/L_d)]$, where $L_d$ is the thickness of the damping layer.  On the inflow and outflow boundaries, the strength function is quadratic in distance, $f(d) = [(1-d)/L_d]^2$.

Sensitivity analysis conducted in this study indicates that the performance of this layer is highly dependent on the relationship between thickness and strength. We found that a shallow layer, $L_d \leq 5~\mathrm{km}$, with a small damping time scale, $\tau < 10$ s, causes the damping region to be somewhat reflective, which is undesirable. Conversely, a time scale that is too large, $\tau > \mathcal{O}(10^{4})$ s, is insufficient to absorb reflected wave motions. An intermediate configuration, using a 10-km thick layer and $\tau = 100$ s, is effective in mitigating spurious wave reflections.

\section{Design of Numerical Experiments}
\label{sec:designOfExperiments}

To design our numerical experiments, we use the Buckingham $\Pi$ theorem \citep{bertrand1878homogeneite} to reduce the number of variables governing AGW characteristics. The variables considered include the flow velocity $U [XT^{-1}]$, buoyancy frequency $N [T^{-1}]$, hill height $h [X]$, and half width of the hill $L [X]$, with fundamental dimensions of length $[X]$ and time $[T]$. Given four variables $(m=4)$ and two fundamental dimensions $(n=2)$, the Buckingham $\Pi$ theorem indicates that the system can be reduced to two ($m-n=2$) independent dimensionless groups. By selecting $U$ and $L$ as repeating variables, we obtain two $\Pi$ groups. The first $\Pi$ group is the hill's shape parameter
\begin{equation}
    Sh = \frac{h}{L},
\end{equation}
which characterizes the geometric steepness of the hill. The second $\Pi$ is the Froude number
\begin{equation}
    Fr = \frac{U}{N L},
\end{equation}
which represents the ratio of inertial to buoyancy forces \citep{Leo_2016, Baines2022} and where
\begin{equation}
    N = \sqrt{\frac{|g|}{\theta_0}\frac{\partial \theta}{\partial z}}
\end{equation}
is the buoyancy frequency, also known as the Brunt-V\"{a}is\"{a}l\"{a} frequency, discussed in detail in Section \ref{sec:oscillatoryNature}, in which $\theta_0=300$ K is the reference temperature and $\partial \theta / \partial z$ is lapse rate (the gradient of potential temperature) along the vertical direction. This dimensional-analysis approach is proposed by \citet{Mehtab2024b} for accurately simulating wind farm-induced AGWs. 

Table~\ref{tab:search_space} describes the combinations of \textit{Fr} and \textit{Sh} parameters that we study, as well as their corresponding atmospheric characteristics. The range of \textit{Fr} from 0.2--0.4 is selected based on practical conditions for large wind farms \citep{smith_gravity_2009,allaerts_gravity_2018}. The incoming velocity is limited to values between the cut-in and cut-out wind speed, 5--25 m/s, for modern utility-scale wind turbines \citep{Markou2009,Huo_2021}, and lapse rate values are constrained to 1--10 K/km that are typical of the free-atmosphere \citep{Sorbjan1996, Abkar2013}. Values of  \textit{Sh} from  0.06--0.10 can be considered analogous to large wind farms, where $h$ corresponds to the rotor tip height and $L$ represents the streamwise length of the wind farm. We used a grid search algorithm to find the combinations of $U$, $N$, and $L$ that yield the desired \textit{Fr} and \textit{Sh} values (Columns 1 and 2 of Table~\ref{tab:search_space}).

\begin{table}
\begin{center}
\def~{\hphantom{0}}
\begin{tabular}{ccccccccc}
~\textit{Fr}~ & ~\textit{Sh}~  & ~$U$(m/s)~ & ~~$N$ (1/s)~~ & ~~$\partial \theta / \partial z$ (K/m)~~ & ~$h$ (m)~ & ~$L$ (m)~ & ~$2\pi L$ (m)~ & ~$2\pi U/N$ (m)~ \\ 
0.2  &  0.06 & ~6  & 0.012~~~  & 0.0044~ & 150  & 2500 &  15707.96 & 3141.59\\ 
0.2  &  0.08 & ~6  & 0.016~~~  & 0.00783 & 150  & 1875 &  11780.97 & 2356.19\\ 
0.2  &  0.10 & ~5  & 0.01667~  & 0.00849 & 150  & 1500 &  ~9424.78 & 1884.96\\ 
0.3  &  0.06 & ~8  & 0.010667  & 0.00348 & 150  & 2500 &  15707.96 & 4712.39\\
0.3  &  0.08 & ~8  & 0.014222  & 0.00619 & 150  & 1875 &  11780.97 & 3534.29\\
0.3  &  0.10 & ~8  & 0.017778  & 0.00967 & 150  & 1500 &  ~9424.78 & 2827.43\\ 
0.4  &  0.06 & 12  & 0.012~~~  & 0.00440 & 150  & 2500 &  15707.96 & 6283.19\\ 
0.4  &  0.08 & 10  & 0.013333  & 0.00544 & 150  & 1875 &  11780.97 & 4712.39\\ 
0.4  &  0.10 & 10  & 0.016667  & 0.00849 & 150  & 1500 &  ~9424.78 & 3769.91\\ 
\end{tabular}
\caption{Values of Froude number (\textit{Fr}), shape factor (\textit{Sh}), incoming velocity ($U$), buoyancy frequency ($N$), hill height ($h$), half length of the hill ($L$), $2 \pi L$ and the buoyancy length scale ($2\pi U/N$) combinations evaluated in the present study.}
\label{tab:search_space}
\end{center}
\end{table}

\section{Wave Quantities of Interest}
\label{sec:qois}

\begin{figure}
\begin{center}
\includegraphics[width=\textwidth]{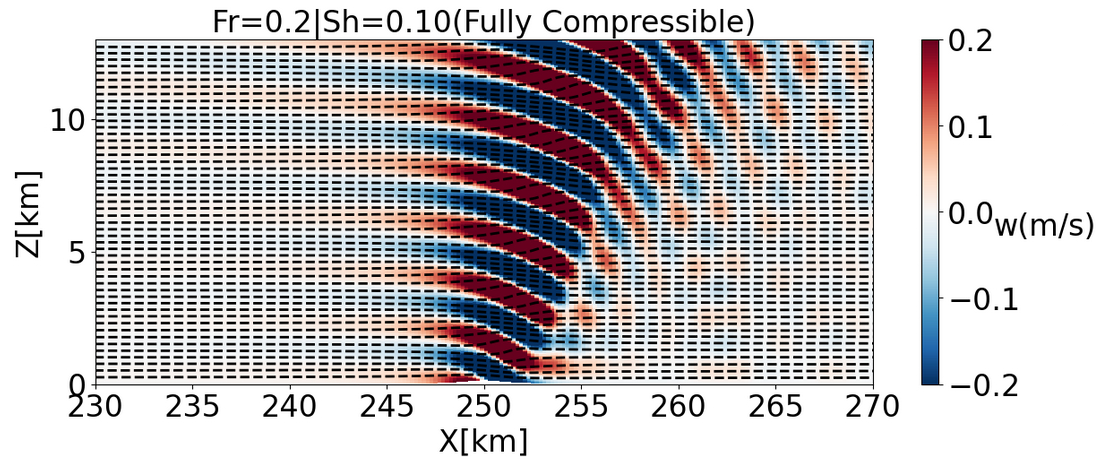}
\caption{AGW amplitudes (defined by pathline displacement) calculated from the vertical motion of air parcels traversing the wave field. Dotted lines track the trajectories of air parcels in the background flow at various heights as they advect through the wavefield in the domain. The maximum vertical displacement of each parcel near the hill determines its respective wave amplitude. The filled contours in the background depict the 2D vertical velocity field.} 
\label{fig:amp_example}
\end{center}
\end{figure}

The three wave characteristics we investigate are amplitude, the wave angle and apparent horizontal and absolute wavelengths. Amplitude, in the context of lee waves, is defined as the vertical displacement of streamlines, as proposed by \citet{Corby1956}. In a stationary wave field like the one in our case, streamlines and pathlines follow the same trajectory. Therefore, if we trace the trajectory of neutrally buoyant Lagrangian particles released at various heights above the hill as they move through the wave field (pathlines), the vertical displacement of each particle is the same as the streamline displacement (see Figure~\ref{fig:amp_example}).  
At each height, we track the vertical displacement of the particles in the vicinity of the hill.  The differences between the mean and maximum displacements across all heights are then noted as the amplitudes of the atmospheric gravity wave field.  In Figure~\ref{fig:amp_example}, we plot particle paths with dashed lines over a representative flow field over a hill. 

The wave angle ($\beta$), crucial for understanding the propagation of wave packets relative to the horizontal flow, is classically defined in linear wave theory using the vertical and horizontal wave numbers: 
\begin{equation}
\beta = \cos^{-1}\left(\frac{k}{\sqrt{k^2+m^2}}\right).
\end{equation}
Note that this relation is based on a single pair of wavenumbers, $k$ and $m$. In reality, however, a hill excites a spectrum of waves, corresponding to a range of $k$ and $m$ values for a given wavelength. Relying on a single pair of wavenumbers to quantify $\beta$ is therefore insufficient to represent the entire wave spectrum.

\begin{figure}
\begin{center}
\includegraphics[width=\textwidth]{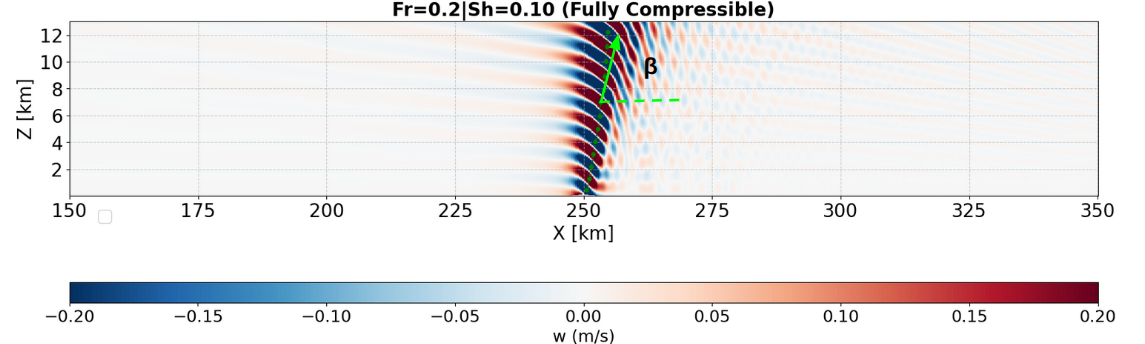}
\caption{An illustration of the calculation of effective flow angles ($\beta$) for one of the cases investigated in this study. It is calculated as the local slope of the upward-facing parabola fitted through the centers of maximas and minimas of the vertical velocity contours at each height.}
\label{fig:beta_phi_calc}
\end{center}
\end{figure}

To address this, we extend the approach by identifying the maxima and minima of the entire wave field and fitting an upward-opening parabola through these points as shown in Figure~\ref{fig:beta_phi_calc}). The local wave angle, $\beta$, is then determined from the slope of the fitted parabola at a given height. This method is inspired by \citet{Queney_1947}.

The last characteristic we examine is the apparent wavelengths. In a 2-D linearized incompressible system, the assumption of a wave-like solution for the vertical velocity field, $\displaystyle e^{kx+mz-\nu t}$, naturally provides single values for wavenumbers $k$ and $m$  \citep{nappo_book,Holton_book,Lin_Book}. However, as air flows over an obstacle in a stably stratified flow, a spectrum (a collection of multiple wavelengths) of waves is excited \citep{nappo_book}. In this study, we estimate the apparent wavelength of the wave spectrum by extracting flow data obtained by each solver to get single values which can be used to optimally design damping layers \citep{Mehtab2024b}. 

We calculate the apparent wavelengths using two different methods. Firstly, we follow an advecting particle in the background flow through the domain; the apparent horizontal wavelength ($\lambda_{X}$) is observed as a variation in the velocity field as it moves through the domain. We do this for parcels at several heights above the hill to get a distribution of apparent horizontal wavelengths with height. The wavelength is measured using a variation of the vertical velocity at each horizontal grid level. This one-dimensional vertical velocity field vector is used for estimating 
the apparent horizontal wavelength based on the autocorrelation function (a method proposed by \cite{Dries_2026}). Since the autocorrelation measures how closely a signal resembles a shifted version of itself, the function will reach its lowest point when the spatial shift corresponds to exactly half a wavelength.  

Consequently, the apparent horizontal wavelength, $\lambda_{X}$, is calculated as double the horizontal displacement that minimizes this autocorrelation
\begin{equation}
    \lambda_a = 2 \times \arg\min_{\xi} r_{ww}(\xi, z),
\end{equation}
where $r_{ww}(\xi, z)$ represents the autocorrelation function defined as
\begin{equation}
    r_{ww}(\xi, z) = \int_{-\infty}^{+\infty} w(x + \xi, z) w(x, z) \, \mathrm{d}x,
\end{equation}
and $\xi$ denotes the horizontal shift.

\begin{figure}
\begin{center}
\includegraphics[width=\textwidth]{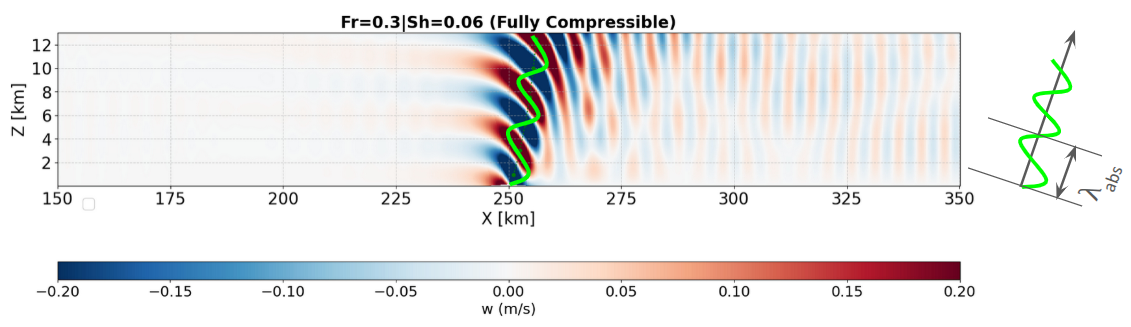}
\caption{The contour map illustrates the vertical velocity field, through which an upward-facing parabola is fitted to align with the centers of negative and positive velocity. Vertical velocity values are sampled along the arc length of this parabola to determine the apparent wavelength. To define the direction of propagation, the slope of the tangent to the parabola is identified as $\beta$, which changes with height.}
\label{fig:abs_wavelengths}
\end{center}
\end{figure}

Secondly, we measure the absolute apparent wavelength ($\lambda_{abs}$) perpendicular to the wavefront along a parabola that is fitted through the centers of the maximas and minimas of the wavefield above the terrain (dots on the red line in Figure~\ref{fig:beta_phi_calc} are the centers of the maxima and minima). The vertical velocity field values are extracted along the arc length of the parabola, and the absolute apparent wavelength is calculated using the autocorrelation method (see Figure~\ref{fig:abs_wavelengths}).


\section{Results and Discussion}
\label{sec:results}

\subsection{Vertical Velocity Fields}
\label{subsec:w_fields}
 

Atmospheric gravity waves induce vertical oscillations in the air and can be analyzed using either horizontal or vertical velocity fields. However, these fields are related by a phase shift, meaning the maxima of the vertical velocity field coincide with the minima of the horizontal velocity fields, and vice versa. For this investigation, we use the vertical velocity field to study the characteristics of the gravity waves. 

\begin{figure}
\begin{center}
\includegraphics[width=\textwidth]{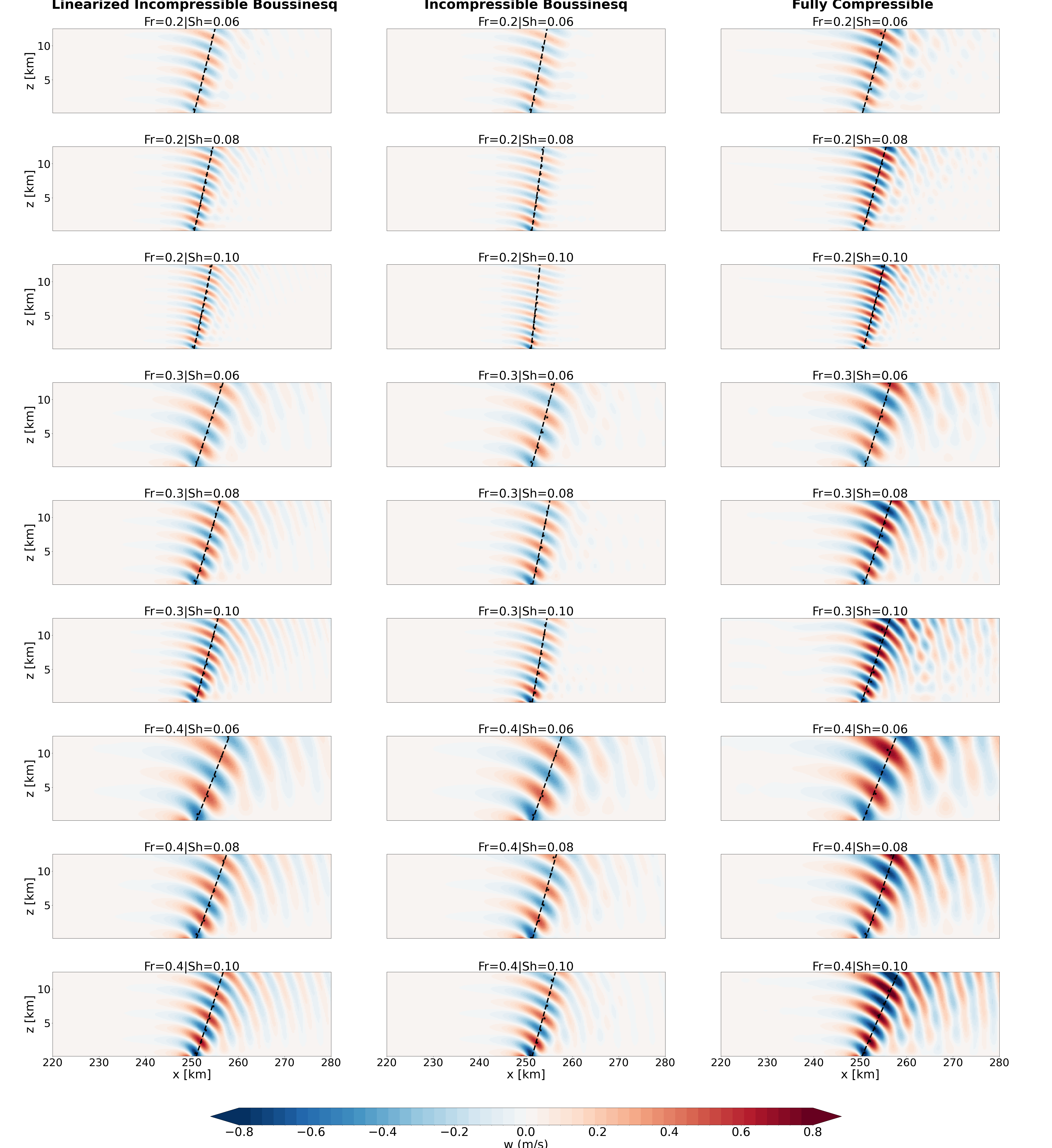}
\caption{Contours of vertical velocity field: (left column) Linearized Incompressible Boussinesq; (middle column) Incompressible Boussinesq; and (right column) Fully Compressible.  Each group of three rows is a different $Fr$, and within each group each row is a different $Sh$, as described in \Cref{tab:search_space}.} 
\label{fig:wContours}
\end{center}
\end{figure}

\begin{figure}
\begin{centering}
\includegraphics[width=\textwidth]{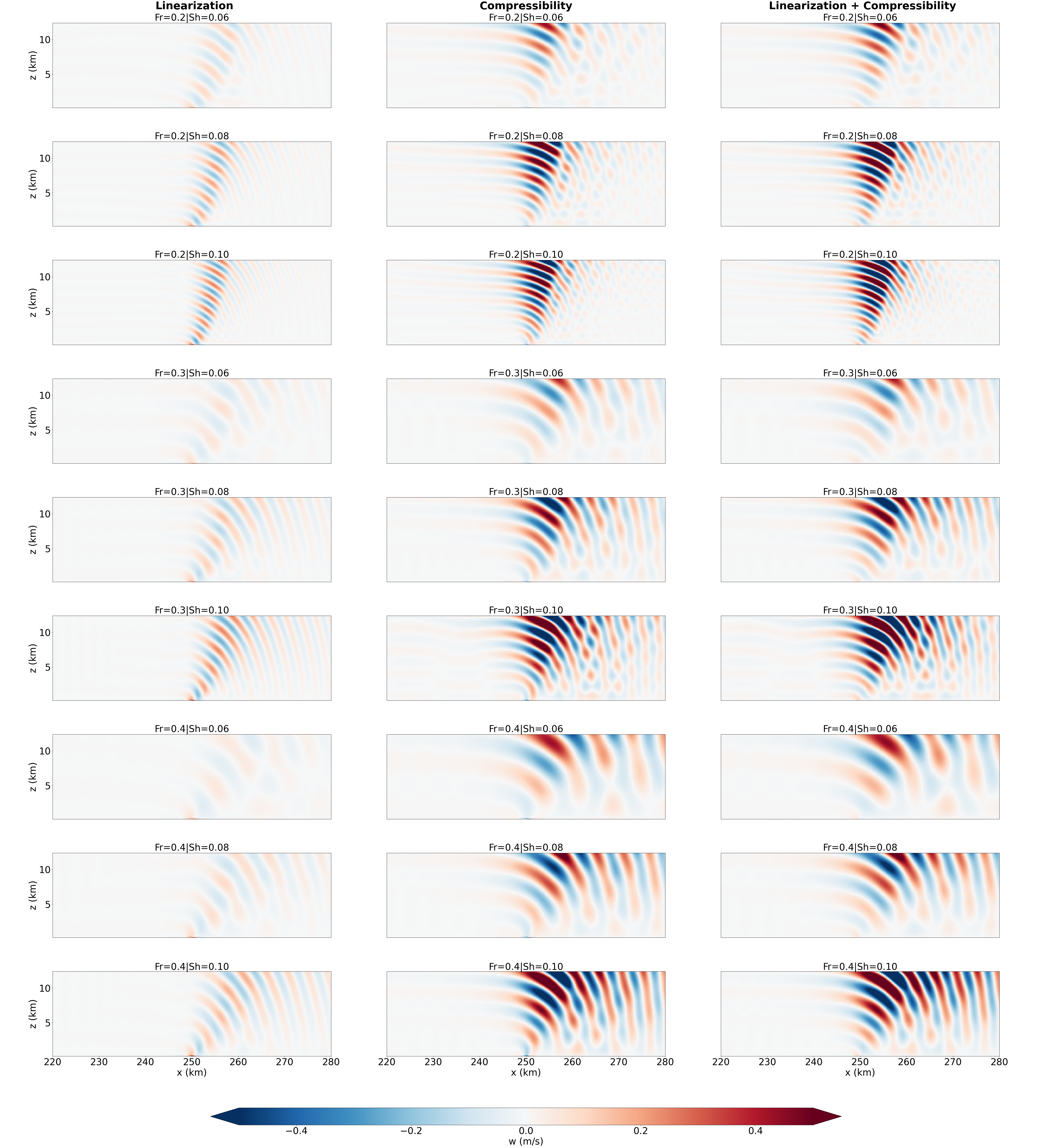}
\caption{Contour plots of the differences in vertical velocity field: (left column) effect of linearization---incompressible Boussinesq minus linearized incompressible Boussinesq; (middle column) effect of compressibility---compressible minus incompressible Boussinesq; and (right column) effect of linearization and compressibility combined---compressible minus linearized incompressible Boussinesq.  Each group of three rows is a different $Fr$, and within each group each row is a different $Sh$, as described in \Cref{tab:search_space}.}
\label{fig:wContours_diff}
\end{centering}
\end{figure}

\begin{figure}
\begin{centering}
\includegraphics[width=1\textwidth]{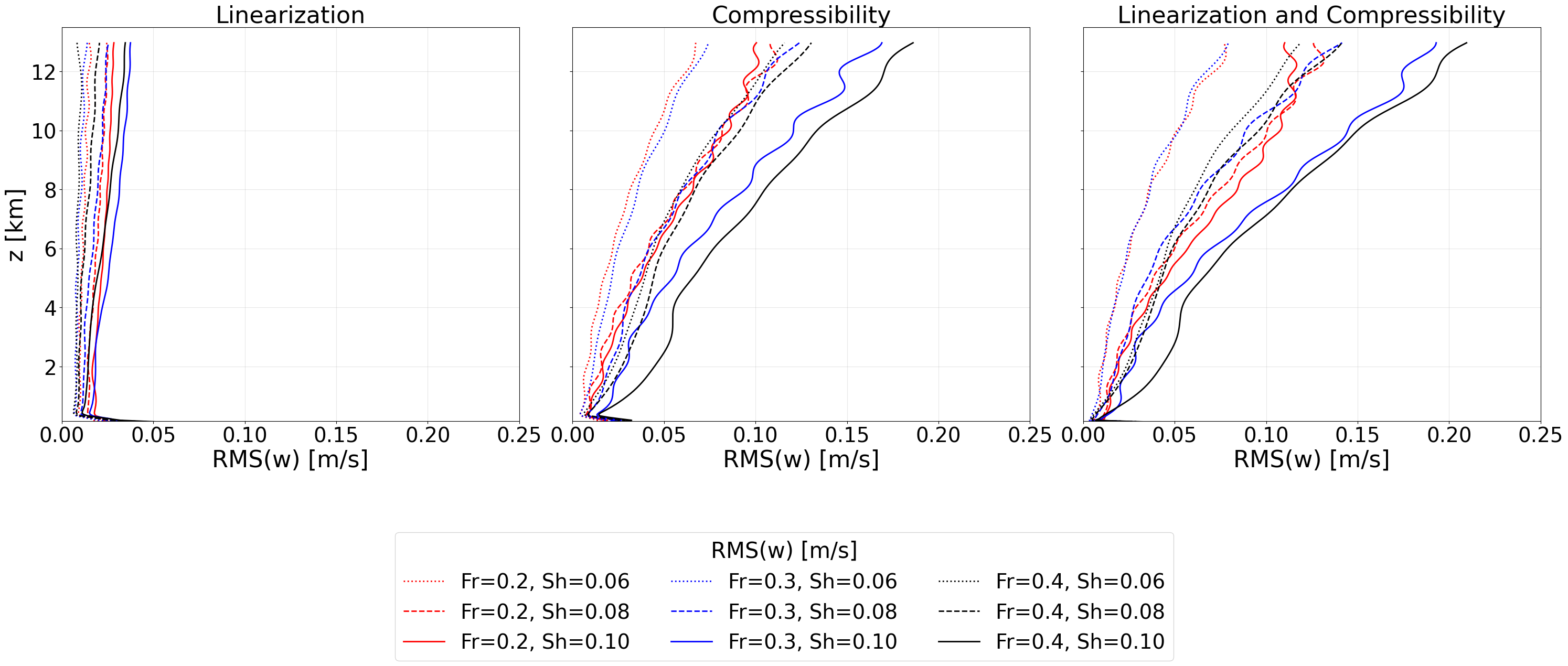}
\caption{Vertical distribution of the root mean square (RMS) difference in vertical velocity field among the different solvers showing the effects of - linearization (SOWFA$-$LBoW) shown in the left column, compressibility (ERF$-$SOWFA) shown in the middle column, and linearization and compressibility (ERF$-$LBoW) shown in the right column.}
\label{fig:wContours_diff_RMS}
\end{centering}
\end{figure}

Figure~\ref{fig:wContours} presents the vertical velocity fields for three forms of the governing equations, and the differences between the models are shown in Figure~\ref{fig:wContours_diff}. We have only plotted the solution 30 km upstream and downstream of the hill to focus on the region of interest. We aggregate the results shown in Figure~\ref{fig:wContours_diff} by taking the root-mean-square (RMS) of the difference in vertical velocity fields predicted by the different equation sets; we do this at each grid level.  Those aggregate results are shown in Figure~\ref{fig:wContours_diff_RMS}.

From the results shown in Figures~\ref{fig:wContours}--\ref{fig:wContours_diff_RMS}, there are a few important observations:
\begin{enumerate}
    \item Using the fully compressible equations to simulate flow over the hill results in notably greater magnitudes of vertical velocity than those predicted by linearized Boussinesq equations (Figure~\ref{fig:wContours}). The impact of linearization, compressibility, and both of these together has been shown in Figure~\ref{fig:wContours_diff}. Note that the contours get darker from the left column to the right column. The vertical distribution of the RMS of the differences in vertical velocity fields predicted by different equation sets shown in Figure \ref{fig:wContours_diff_RMS} confirms this trend, while also showing that compressibility effects become more pronounced at higher altitudes and for higher slopes of the hill. Consequently, air parcels experience greater vertical displacement amplitudes while traversing the wave field in these regions (Figure \ref{fig:amplitudes}).
    \item The shape parameter (\textit{Sh}) describes the hill slope because it is the ratio of hill height to half-height length. According to linear theory, hill width dictates the apparent wavelength of the AGWs and hill height dictates their amplitude. With the hill height fixed at 150 m, increasing $Sh = h/L$ translates to reducing the half height length ($L$), resulting in shorter horizontal and vertical wavelengths. This shift is evidenced in Figure~\ref{fig:wContours} by the increased number of vertical velocity extrema. In Figure~\ref{fig:wContours_diff}, the difference contours for a constant Froude number ($Fr$) intensify as $Sh$ increases, a trend further quantified by the vertical RMS profiles in Figure~\ref{fig:wContours_diff_RMS}. Across all three images in Figure~\ref{fig:wContours_diff_RMS}, higher $Sh$ values correlate with increased RMS errors that grow with altitude. Notably, when compressibility is incorporated, these errors become even more pronounced at greater heights. This suggests that for steeper hills, a compressible flow solver is essential for accurately capturing AGW characteristics in the middle atmosphere. Ultimately, these results demonstrate that the shape parameter significantly influences both the apparent wavelength and the magnitude of the vertical velocity field.
    \item The Froude number, \textit{Fr} $=U/NL$, represents the ratio of inertial to buoyancy forces. Increasing \textit{Fr} involves varying $U$ and $N$ while keeping $L$ fixed for a given $h$ (see Table~\ref{tab:search_space}). Thus, the increase in \textit{Fr} increases $U/N$ and the buoyancy length scale, $2\pi U/N$, resulting in longer vertical wavelengths as observed in Figure~\ref{fig:wContours}. This is shown by a reduction in the number of distinct blue and red regions at higher \textit{Fr} values. Additionally, as \textit{Fr} increases and inertial forces begin to dominate, the gravity wave angle relative to the horizontal decreases. This is evidenced by the shallower inclination of the dotted lines in the bottom three rows compared to the upper rows, a behavior also highlighted by \cite{Dries_2026}. However, the degree to which compressibility affects the wave angles is slightly more nuanced and will be discussed in \cref{subsec:directions}.

\end{enumerate}

\subsection{Wave Amplitudes}
\label{subsec:disp}

\begin{figure}
\begin{center}
\includegraphics[width=0.85\textwidth]{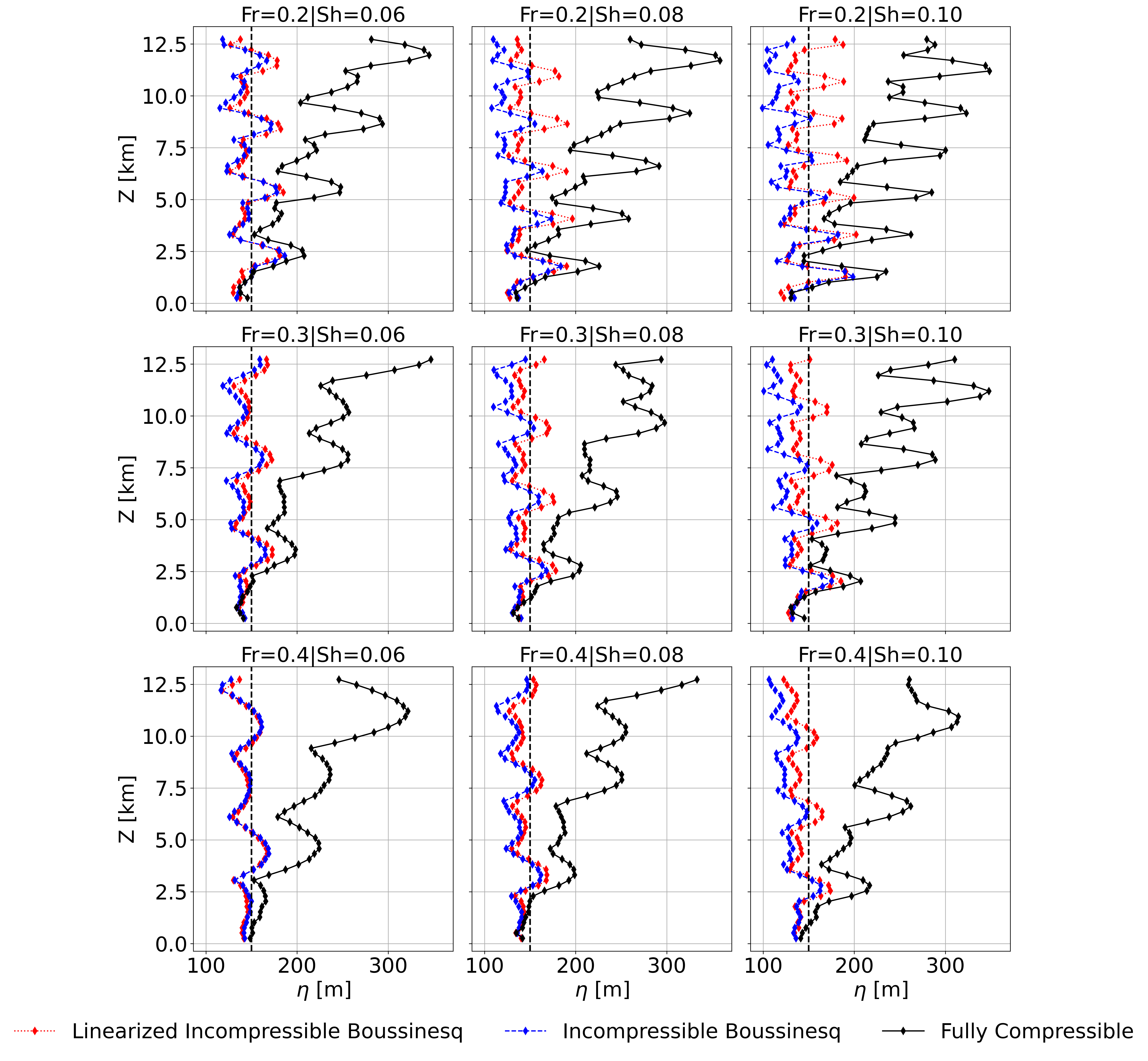}
\caption{Vertical displacements of pathlines at various heights above the hill. The displacements for each parcel in the vicinity of the hill are used to develop this plot (see Figure~\ref{fig:amp_example}).  The dashed black line in each plot is the hill height, which is 150 m in all cases.} 
\label{fig:amplitudes}
\end{center}
\end{figure}

Previous studies, such as \cite{allaerts_boundary-layer_2017} and \cite{Lanzilao_2024} used incompressible Boussinesq LES to simulate flow over wind farms and predicted air parcel displacements up to 100 meters, which is on the order of the hub height of many wind farms.  Figure~\ref{fig:amplitudes} shows the magnitude of amplitudes calculated as streamline displacements of neutrally buoyant particles as they traverse the stationary gravity wave field at various heights above the hill.  In Figure~\ref{fig:amplitudes}, the incompressible Boussinesq results (which are most like what those studies used) show similar trends in results: displacements roughly equal to the height of the perturbation itself 150 m (shown by the dashed black line in each of the plots in Figure \ref{fig:amplitudes}).  We see similar results for linearized incompressible Boussinesq flow.  However, we see substantially higher displacements at higher altitudes in the fully compressible model.

\begin{figure}
\begin{center}
\includegraphics[width=0.95\textwidth]{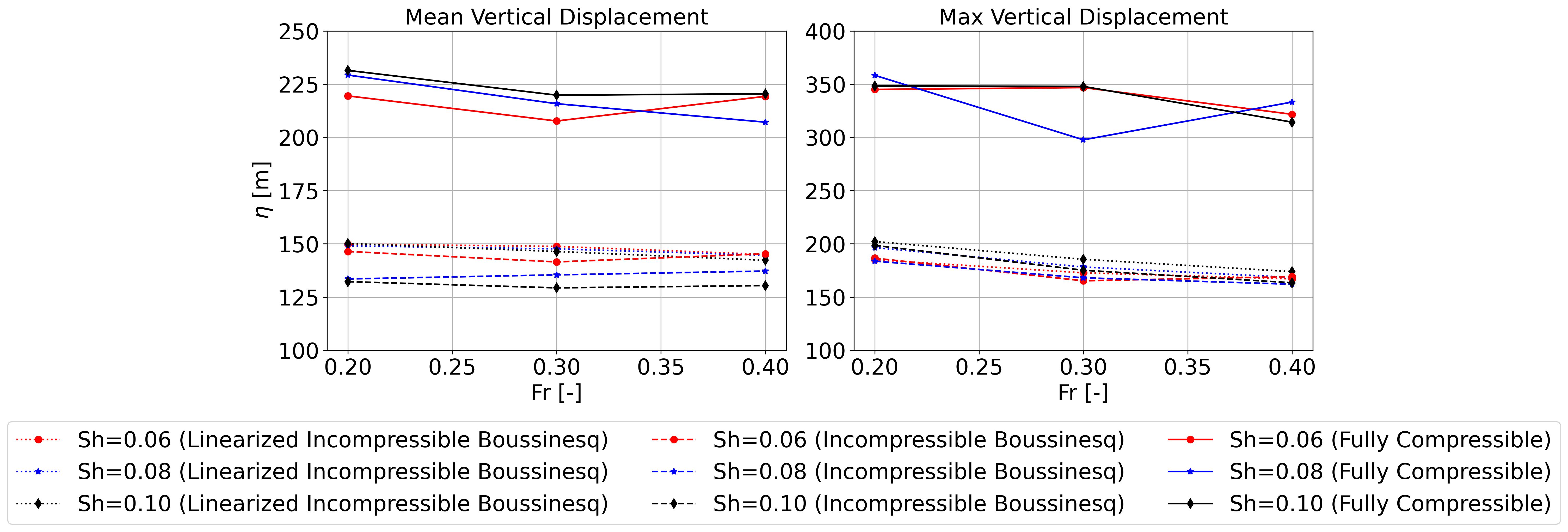}
\caption{Maximum and mean of all vertical displacements of pathlines at various heights above the hill.} 
\label{fig:amplitudes_mean_max}
\end{center}
\end{figure}

Consider an air parcel embedded in a uniform background flow. The parcel experiences vertical displacements as it traverses a stationary gravity wave field generated by a hill in a uniformly stratified atmosphere. It ascends in regions of positive vertical velocity perturbation ($w' > 0$) and descends in regions of negative perturbation ($w' < 0$).  When displaced upward, the parcel expands, cools, and becomes negatively buoyant. The parcel undergoes vertical oscillations about its equilibrium position. The two main forces that control this oscillatory motion are the vertical pressure gradient force that tends to push the parcel upward and the restoring negative buoyancy force that tries to push it back down. In a Boussinesq framework, the restoring buoyancy acts like a linear spring as elucidated in \cref{sec:oscillatoryNature}. In contrast, for a compressible framework, the buoyancy force acts like a non-linear spring in that the spring constant reduces with altitude because z appears in the denominator of equation~\ref{eq:buoy_freq_comp}. This means that at higher altitudes, the tendency of the air parcel to oscillate with greater amplitude is higher (due to a smaller restoring force and higher magnitudes of vertical velocity extrema). This is confirmed in Figure~\ref{fig:amplitudes}.

The structure of $w'(x,z)$ is determined by the hill geometry, the atmospheric stability and the approximations made in governing flow equations. This is a two-dimensional wave field exhibiting a wave-like character both horizontally and vertically. If an air parcel is suspended at the height of the phase change (directly between the maxima and minima), the amplitude of the oscillation is minimized. This is more pronounced for low \textit{Fr} cases where the wave train propagates up more than downstream and is stacked more directly over the hill. This happens less for high $Fr$ in which the wave train is arranged more in concentric arcs.

Further, Figure~\ref{fig:amplitudes_mean_max} shows the mean and maximum values of amplitudes calculated from the distribution of the vertical displacement of pathlines. It can be observed that the fully compressible code shows higher values of mean and maximum pathline displacements than its linearized incompressible Boussinesq and incompressible Boussinesq counterparts. This can be attributed to the higher displacements of the pathlines at higher altitudes, as discussed earlier.

For wind-farm-induced gravity waves, the amplitude determines the extent of vertical oscillations in the incoming air. Understanding these amplitudes helps clarify how gravity waves affect the wind resource of a leeward wind farm. Since wind farms typically feature turbines having hub heights that are a few hundred meters tall and the farm spans several kilometers, they have a low shape factor ($Sh$). As shown in Figure~\ref{fig:amplitudes}, at low $Sh$ and altitudes below 1 km, the amplitudes predicted by the incompressible Boussinesq solvers and the fully compressible framework are in close agreement.

\subsection{Wave Angles}
\label{subsec:directions}

Recent studies have reported that operational wind farms often produce less energy than initially estimated during the siting phase \citep{Bleeg2018}. This discrepancy has been attributed to \textit{wind farm blockage}, a phenomenon characterized by the development of an adverse pressure gradient upstream of the farm, leading to flow deceleration before the first row of turbines \citep{Bleeg2018}. This mechanism bears semblance to the formation of AGWs in stably stratified flows over wind farms (that act like porous hills), where adverse pressure gradients ahead of a wind farm generate vertically propagating wave motions \citep{Allaerts_thesis, allaerts_boundary-layer_2017}. Given this similarity, it is essential to investigate how wind farm characteristics (such as turbine height) and atmospheric stratification influence the generation and propagation of AGWs, particularly the angle at which these waves propagate relative to the vertical. The wave angle $\beta$ determines the orientation of AGW wave packets, alternating regions of positive and negative vertical velocity perturbations ($+w'$ and $-w'$). This in turn governs zones of flow acceleration and deceleration within the wind farm, especially between turbine rows, and in the lee of the farm.


The vertical tilt of the wavefield ($\beta$) determines the direction in which the gravity waves propagate with respect to the horizontal. $N$ determines the vertical propagating speed of the waves, while the horizontal advection is determined by the background wind speed $U$. Changing $U$ and $N$ determines the direction of the wave propagation. Increasing U while keeping N fixed (translates to an increasing $Fr$) causes the waves to be advected more downstream.  This is observed in Figure~\ref{fig:wContours}, where in the latter rows (cases with high \textit{Fr}), the wave packets bend more downwind, as was also reported by \cite{Dries_2026}. 

\begin{figure}
        \centering
        \includegraphics[width=\textwidth]{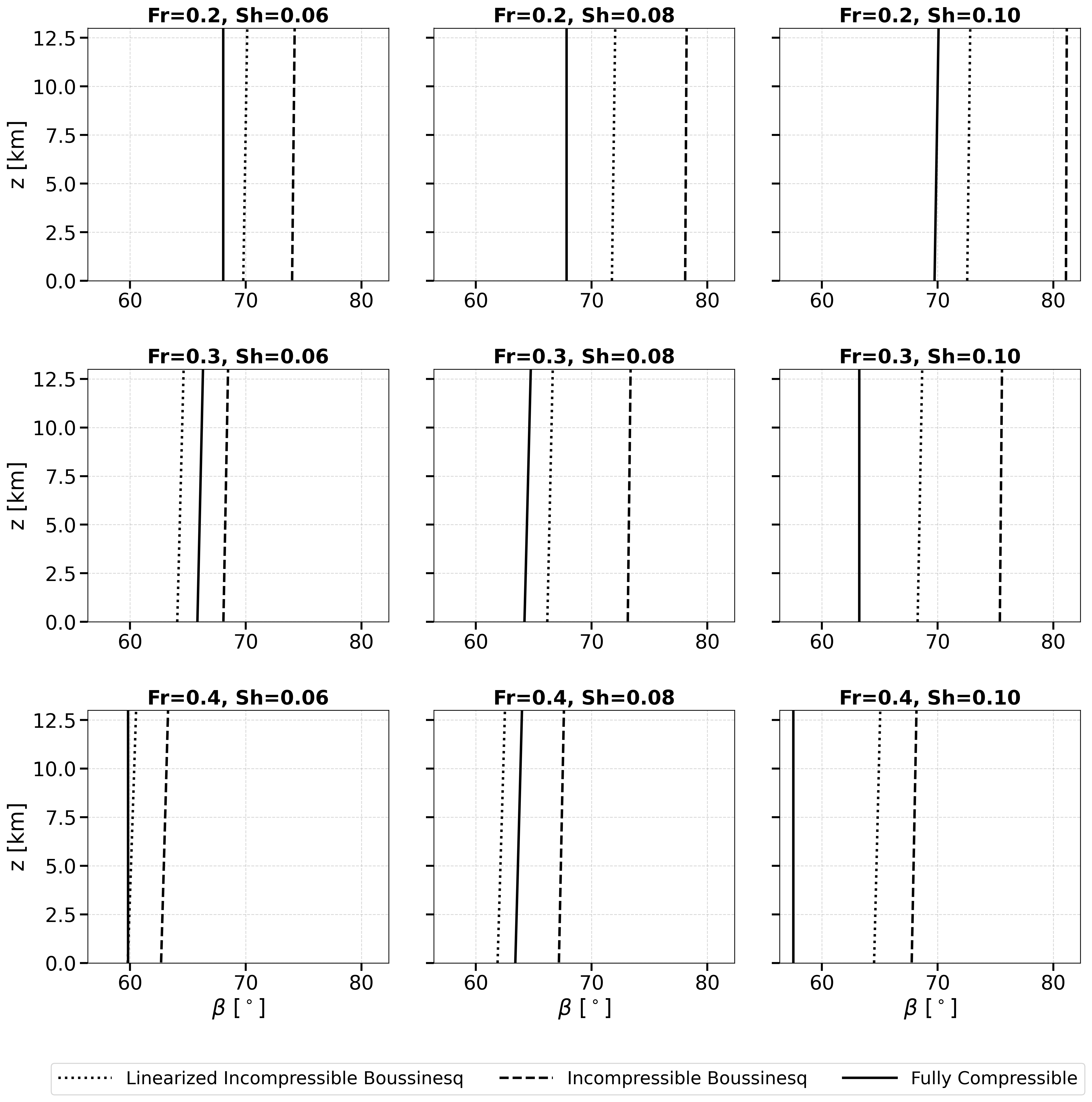}

    \caption{Vertical distribution of wave angles—calculated from the flow fields as the tilt of upward-facing parabolas (Section~\ref{sec:qois})—for the linearized incompressible Boussinesq, incompressible Boussinesq, and fully compressible cases.}
    \label{fig:wave_angles}
\end{figure}

Figure~\ref{fig:wave_angles} shows the vertical distribution of wave angles measured as the slope of the fitted parabola at heights above the hill. This too shows similar trends, where, comparing the bottom row to the rows above, we see a decrease in $\beta$ with increasing \textit{Fr}.  For flow regimes having low \textit{Fr} and high \textit{Sh} (highly sloped hills), we see a higher value of $\beta$, which implies that the lower kinetic energy of the incoming flow is insufficient for the air parcel to go over the hill. Instead, it causes a flow field that is more vertically oriented. This decreases with an increase in \textit{Fr}. This is confirmed in Figure~\ref{fig:wave_angles} and also upon a close perusal of Figure~\ref{fig:wContours}. This implies that there may be a significant difference in which row of turbines within a large farm experiences local acceleration (subsequently higher wind power output) and vice versa, depending on which tool is used to simulate and study flows around wind farms.


Further, for the compressible cases, at higher altitudes, we see the effect of $N$, which creates a \textit{Fr} variation with height. Lower values of $N$ cause the \textit{Fr} to be higher at these altitudes, which causes the gravity waves to propagate at lower $\beta$. This can be observed in Figure~\ref{fig:wContours}, where in the compressible case (right column). This downwind propagation influences the flow in the wake of the hill (or wind farm), making the distance downstream of the hill a critical factor, whereas using an incompressible framework would miss this important influence on the wind farm. Thus, the AGW propagation angle emerges as a critical parameter in the accurate siting and performance forecasting of wind farms.

\subsection{Apparent Wavelengths}
\label{subsec:wavelengths} 

According to linear theory, horizontal wavelength scales with a factor of $2\pi L$, where $L$ is the hill's half-length (a parameter determining the shape of the hill) \citep{nappo_book}. Vertical wavelength, on the other hand, scales with $2\pi U/N$, where $N$ is the buoyancy frequency and $U$ is the background velocity \citep{nappo_book}. In the case of compressible flow, we have shown that the buoyancy frequency varies with height; this implies that we cannot use a constant \textit{Fr} to determine the scaling parameter. Due to this, we have chosen to normalize the apparent wavelengths by the half length of the hill (L).


\begin{figure}
    
        \centering
        \includegraphics[width=0.95\textwidth]{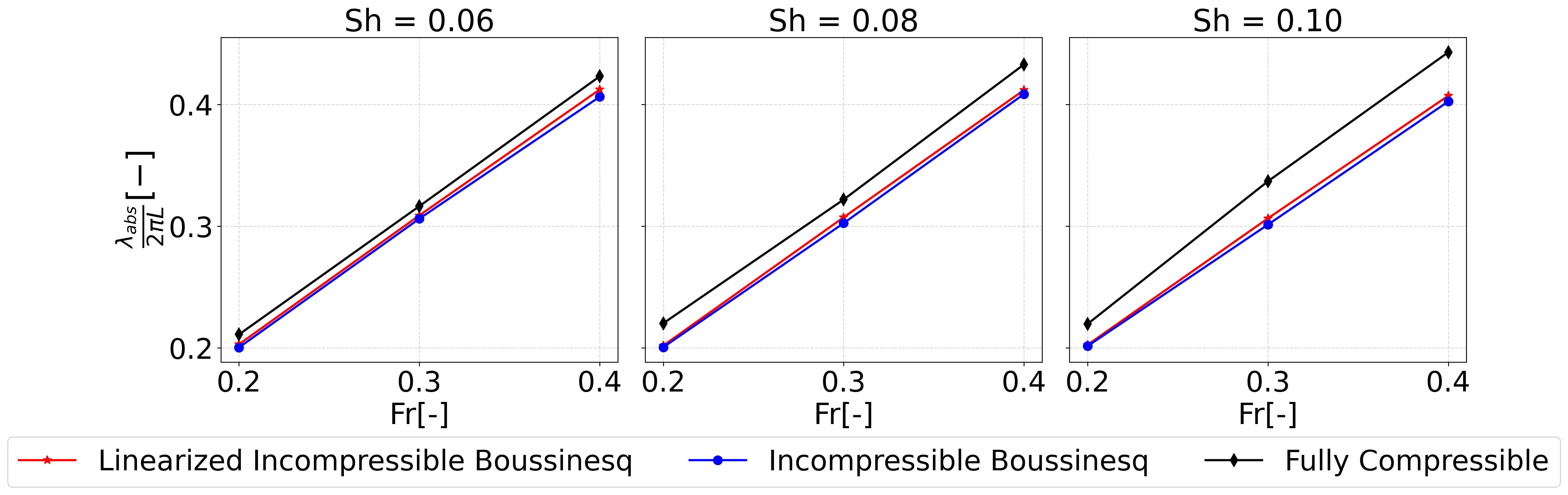}

    \caption{Variation of non-dimensional apparent absolute wavelength with \textit{Fr} for a given \textit{Sh} for the linearized incompressible Boussinesq, incompressible Boussinesq and fully compressible framework}
    \label{fig:app_abs}
\end{figure}

Figure \ref{fig:app_abs} illustrates the variation of the normalized apparent absolute wavelength as a function of the Froude number ($Fr$). Notably, even within the fully nonlinear incompressible Boussinesq framework, the wavelength scales nearly perfectly with $2\pi U/N$, demonstrating a remarkable agreement between theoretical predictions from linear wave theory and the numerical computations from the three different forms of the governing equations. Overall, the results show a consistent trend where the absolute wavelength increases with $Fr$. This behavior is expected, as an increase in $Fr$ corresponds to a larger buoyancy length scale ($2\pi U/N$), which then shows up as larger apparent absolute wavelengths. Furthermore, the apparent absolute wavelengths are highly similar between the linearized and the incompressible Boussinesq frameworks. However, notable deviations emerge within the compressible framework; these discrepancies amplify both as the hill slope ($Sh$) steepens and as $Fr$ increases. 

\begin{figure}
    
        \centering
        \includegraphics[width=\textwidth]{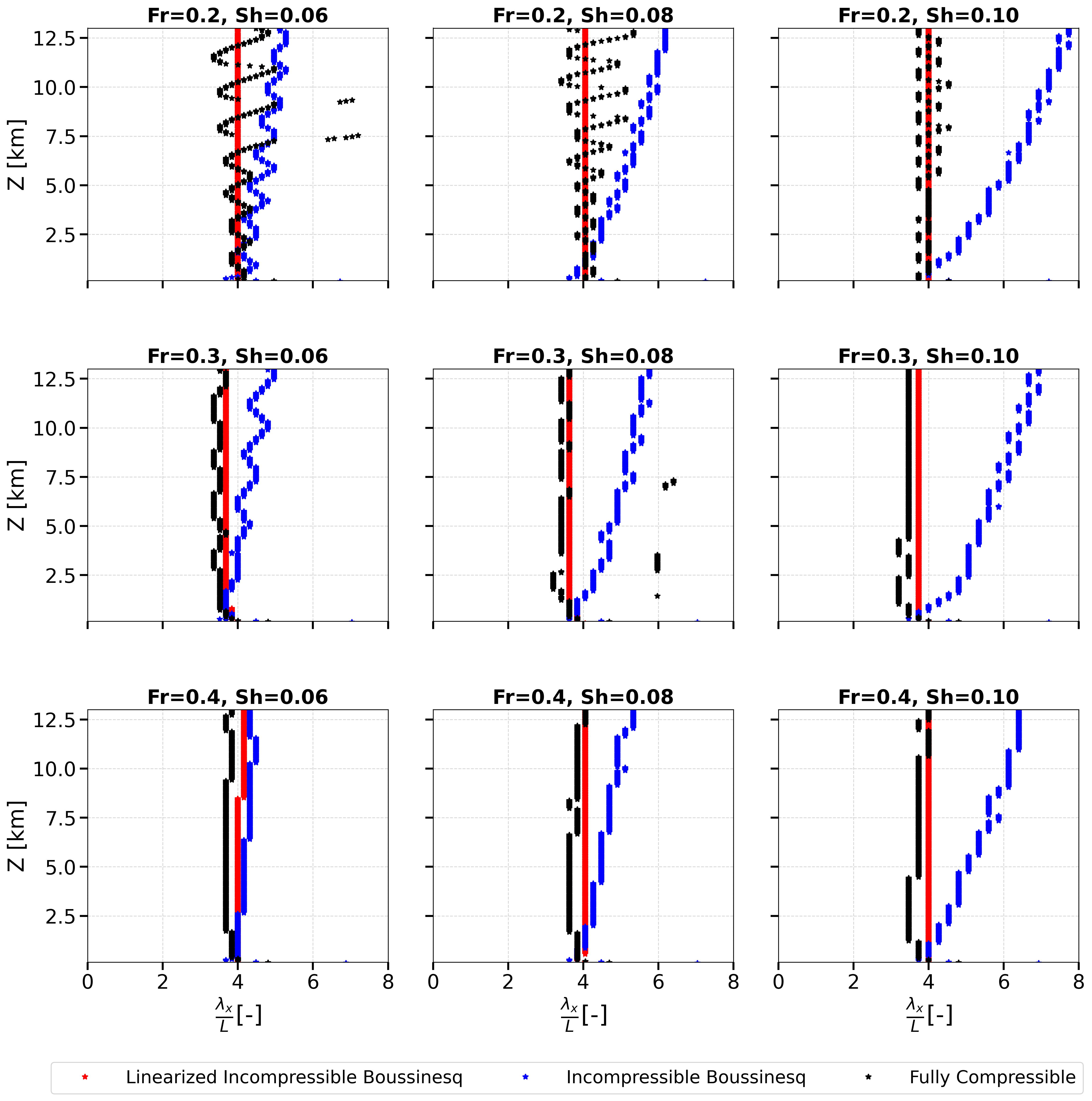}

    \caption{Vertical distribution of apparent horizontal wavelengths normalized by the half length of the hill ($L$)for a linearized incompressible Boussinesq, incompressible Boussinesq and fully compressible system}
    \label{fig:app_X}
\end{figure}

Figure \ref{fig:app_X} illustrates how the normalized apparent horizontal wavelength varies with height across different frameworks. For the linearized incompressible Boussinesq framework, this value remains almost constant with height (centered around 4) for all simulated cases, a trend also observed by \cite{Dries_2026}. However, the incompressible Boussinesq framework exhibits a consistent increase in wavelength with height, meaning the waves become longer with height. This increase of apparent horizontal wavelength with height becomes even more pronounced over steeper hill slopes ($Sh$). This can be explained by closely examining the first column of figure~\ref{fig:wContours_diff}. For a given \textit{Fr} as \textit{Sh} increases, there are larger differences in the vertical velocity field between the linearized incompressible Boussinesq framework and the incompressible Boussinesq framework in that the differences are spread out over a larger horizontal distance downstream of the hill. These differences in the spatial differences in the vertical velocity field could be the reason why we see a growth in the apparent horizontal wavelength with height for the incompressible Boussinesq framework. 

Finally, in the compressible framework, wavelengths vary significantly at low Froude numbers (\textit{Fr}), as seen in the top row of Figure \ref{fig:app_X}. At low \textit{Fr}, buoyancy forces dominate inertial forces, resulting in a weaker restoring force on displaced air parcels. Furthermore, as explained in Section \ref{sec:oscillatoryNature}, the Brunt-Väisälä frequency ($N$) varies with height in a compressible framework, causing a proportional variation in \textit{Fr} with height.

In the context of windfarm-induced internal gravity waves, this implies that as the hub height of the turbines increases and the length of the farm decreases (which corresponds to \textit{Sh}), using a compressible framework could predict different wavelengths as compared to an incompressible Boussinesq framework.

\section{Conclusions}
\label{sec:conclu}
The linearized incompressible Boussinesq set of equations and its associated linear dispersion relation, as well as the nonlinear incompressible Boussinesq set of equations, are widely used to estimate wave characteristics of AGWs \citep{nappo_book,Lin_Book,Holton_book}, such as wave angles, amplitudes, and phase/group velocities. We have presented arguments that there are cases, potentially important ones with real-world consequences, where these simplifications may not be appropriate and can lead to inaccurate estimations of atmospheric gravity wave (AGW) behavior, particularly in buoyancy-dominated or low-Froude-number flows. In particular, many simulations of wind farm-induced AGWs use incompressible Boussinesq LES as their primary tool to quantify impacts on wind power variability within wind farms. The present study (using flow over a hill as an analogue for a wind farm) shows that this choice, while being acceptable for simulating wind farm aerodynamics, can lead to discernible differences under certain flow conditions.

Our study yields several conclusions:
\begin{enumerate}
    \item For the same hill shape and atmospheric stratification, compressible codes generate AGWs that are qualitatively similar to incompressible Boussinesq codes. This consistency increases confidence that wind farm simulations conducted with compressible codes are also likely to generate gravity waves similar to those predicted by incompressible Boussinesq solvers.
    \item For the range of Froude numbers and shape parameters investigated in this study, the compressible model generally predicts higher parcel displacements compared to both the incompressible Boussinesq model and linear theory, as it allows for variation of density. While the incompressible Boussinesq model does capture wind farm–induced AGWs, it can mispredict air parcel displacements, which could lead to inaccurate estimates of energy yields from the turbines. In the case of a real wind farm immersed in an atmospheric boundary layer, the inlet condition could change over time due to the diurnal variation of the surface heat flux, which can excite travelling AGWs that could contribute to changes in power production with time. Unlike a solid hill, which forces the flow entirely over it, a semi-permeable hill allows the flow to pass both through and above the wind farm. This could result in weaker AGW excitation compared to the flow over a solid hill. However, the degree to which porosity (flow through and around wind farms) affects wave amplitudes requires further investigation for the compressible flow solvers.
    \item At low \textit{Fr} (high $N$, low $U$), where the buoyancy forces are stronger than inertia forces, the streamline displacement or wave amplitudes increase with height in the compressible framework. In contrast, they stay approximately the same as the height of the hill for the incompressible Boussinesq framework and the linearized incompressible Boussinesq framework. 
    
    \item Although gravity wave amplitudes predicted by compressible and incompressible Boussinesq solvers closely align at lower altitudes, they diverge significantly at higher altitudes. This discrepancy implies that while Boussinesq codes can accurately evaluate individual wind-farm-induced gravity waves impacts, they are inadequate for broader mesoscale applications. Specifically, simulating flow over multiple wind farms requires an accurate quantification of the incoming kinetic energy at a leeward farm, which depends on the net vertical momentum flux of the gravity waves induced upstream. Incompressible Boussinesq models may not be able to capture this altitude-dependent amplitude variations. Furthermore, because these models neglect this vertical variability, they may potentially lead to inaccurate estimations of the total energy carried away by the gravity waves.
    
    \item The incompressible formulation of $Fr = U/NL$ usually considers $N$ to be dependent only on potential temperature lapse rate. However, we have shown that in a compressible flow scenario (which is a truer representation of the atmosphere), the \textit{N} depends on the potential temperature lapse rate and the neutral height from which the parcel is displaced. Consequently, this vertical variation in $N$ must be accounted for, rendering $Fr$ height-dependent as well. Failing to use this corrected, height-varying formulation can lead to an erroneous quantification of wave angles and their vertical evolution. Furthermore, this variation in $N$ inherently alters the behavior of apparent wavelengths with height under compressible conditions.

    \item Vertical variations in the buoyancy frequency ($N$) alter the propagation angles of wind-farm-induced gravity waves. While this effect may be negligible for an isolated wind farm, it may be of importance due to its influence on the direction and magnitude of wave energy propagation across multiple wind farms. Consequently, using incompressible Boussinesq solvers to simulate multi-farm interactions could introduce errors, as they may fail to capture how a variable $N$ profile redirects the vertical wave field and alters the spatial distribution of regions of localized flow acceleration and deceleration.

    \item While the normalized apparent absolute wavelengths scale linearly with $Fr$ across both numerical approaches, the fully compressible models consistently predict larger apparent horizontal wavelengths than their incompressible Boussinesq counterparts. Under a linearized Boussinesq framework, the normalized apparent horizontal wavelength remains invariant, centered at 4. Conversely, the fully nonlinear incompressible framework exhibits a strong height-dependence at low $Fr$—an effect further amplified by steeper topography. Similarly, the fully compressible framework displays pronounced horizontal wavelength variations at high altitudes for low $Fr$, though this variability diminishes at higher $Fr$. Collectively, these trends imply that to accurately simulate wind farm-induced gravity waves at low $Fr$, numerical boundary treatments—such as upper-level Rayleigh damping layers and sponge regions—must be carefully configured to accommodate these height-dependent wavelength variations.

\end{enumerate}
 
We conclude that using the incompressible Boussinesq equations generally gives similar AGW characteristics to using the compressible equations, but given the potential for large and/or consequential discrepancies, while simulating flow over multiple wind farms, the use of compressible solvers seems worth considering. 

In the present work, we have not considered the anelastic framework. But it can be used to capture the impact of gravity waves without the computational overhead of the compressible framework. A future research direction would be to investigate how using an anelastic flow solver impacts simulated gravity wave characteristics. Nonetheless, applying these hill induced gravity wave findings to wind farm induced gravity waves comes with limitations. Unlike our fully stratified atmosphere, actual wind farms operate within the atmospheric boundary layer (that contains a mixed layer, capped by an inversion and a free atmosphere aloft). While boundary layer flow over topography can trigger both inversion-trapped lee waves and free-atmosphere internal gravity waves, this study models only the latter. While recognizing these caveats, we anticipate that these insights will ultimately support improved wind plant siting and more reliable power predictions through informed solver choices.



\backsection[Acknowledgements]{We dedicate this paper to Dries Allaerts.  He and his work inspire us daily.}

\backsection[Funding]{This work was supported in part by the Indiana University Environmental Resilience Institute and Lilly Endowment, Inc. through its support for the Indiana University Pervasive Technology Institute.  This work was authored in part by the National Laboratory of the Rockies for the U.S. Department of Energy (DOE), operated under Contract No. DE-AC36-08GO28308. Funding provided by the U.S. Department of Energy Office of Critical Minerals and Energy Innovation Integrated Energy Systems Office. The views expressed in the article do not necessarily represent the views of the DOE or the U.S. Government. The U.S. Government retains and the publisher, by accepting the article for publication, acknowledges that the U.S. Government retains a nonexclusive, paid-up, irrevocable, worldwide license to publish or reproduce the published form of this work, or allow others to do so, for U.S. Government purposes.}

\backsection[Declaration of Interests]{The authors report no conflict of interest.}


\backsection[Author ORCIDs]{H. Sivanandan, https://orcid.org/0000-0002-1759-5165, B. Kravitz, https://orcid.org/0000-0001-6318-1150, M. Churchfield,  https://orcid.org/0000-0001-9906-7288, E. Quon, https://orcid.org/0000-0002-8445-5840, M.A. Khan, https://orcid.org/0000-0001-7953-0941}

\backsection[Credit author statement
]{Conceptualization: H.S., M.C., B.K., M.K.; Methodology: H.S.; Software: E.Q., M.C.; Validation: H.S., M.C.; Formal Analysis: H.S.; Investigation: H.S., M.K.; Data Curation: H.S.; Writing: H.S., M.C., B.K.; Reviewing and Editing: H.S., M.C., B.K., E.Q., M.K; Visualization: H.S.; Supervision: M.C., B.K.; Project Administration: B.K.}

\bibliographystyle{jfm}
\bibliography{jfm}

\end{document}